\documentclass[aps,prd,preprint,showpacs,preprintnumbers,nofootinbib,superscriptaddress]{revtex4}
\usepackage{ae}
\usepackage{bm} 
\usepackage{color}
\usepackage{amsmath}
\usepackage{amssymb}
\usepackage{amsfonts}
\usepackage{graphicx}
\usepackage{slashed}
\usepackage{wasysym}

\allowdisplaybreaks

\begin{document}
\setlength{\unitlength}{1mm}
\preprint{TUM-EFT 29/11}
\title{Symmetries of the Three Heavy-Quark System and the Color-Singlet Static Energy at NNLL}
\author{Nora Brambilla}
\affiliation{Physik-Department, Technische Universit\"at M\"unchen,
James-Franck-Str. 1, 85748 Garching, Germany}
\author{Felix Karbstein}
\affiliation{Helmholtz-Institut Jena, Fr\"obelstieg 3, 07743 Jena, Germany}
\affiliation{Theoretisch-Physikalisches Institut, Friedrich-Schiller-Universit\"at Jena, Max-Wien-Platz 1, 07743 Jena, Germany}
\author{Antonio Vairo}
\affiliation{Physik-Department, Technische Universit\"at M\"unchen,
James-Franck-Str. 1, 85748 Garching, Germany}

\date{\today}

\begin{abstract}
We study the symmetries of the three heavy-quark system under exchange of the quark fields 
within the effective field theory framework of potential non-relativistic QCD. 
The symmetries constrain the form of the matching coefficients in the effective theory.
We then focus on the color-singlet sector and determine the so far unknown leading ultrasoft contribution
to the static potential, which is of order $\alpha_{\rm s}^4\ln\mu$, and consequently to the static energy, 
which is of order  $\alpha_{\rm s}^4\ln\alpha_{\rm s}$. Finally, in the case of an equilateral geometry, 
we solve the renormalization group equations and resum the leading ultrasoft logarithms 
for the static potential of three quarks in a color singlet, octet and decuplet representation. 
\end{abstract}


\maketitle

\section{Introduction}
\label{intro}
Bound states of a heavy quark $Q$ and an antiquark $\bar Q$
have been the subject of extensive theoretical studies since the early
days of quantum chromodynamics (QCD). Relatively less attention has been paid to bound states 
of three heavy quarks ($QQQ$), also referred to as triple heavy baryons, 
as a consequence of their still missing experimental evidence.  
Nevertheless there is an ongoing theoretical activity devoted 
to their study mostly driven by lattice computations~\cite{Sommer:1985da,Takahashi:2000te,Takahashi:2002bw,
Suganuma:2000bi,Alexandrou:2002sn,Takahashi:2002it,Takahashi:2003ty,
Bornyakov:2004uv,Bornyakov:2004yg,Takahashi:2004rw,Hubner:2007qh,Iritani:2010mu,Meinel:2012qz}, 
but also by phenomenological analyses (for a review see~\cite{Klempt:2009pi})
and more recently by effective field theory methods~\cite{Brambilla:2005yk,Brambilla:2009cd,Vairo:2010su}.
The theoretical interest is mainly triggered by the geometry of these systems, which allows to address questions 
that are inaccessible with two-body systems. Examples are the minimal energy configuration of three quarks 
in the presence of a confining potential or the origin of a three-body interaction.
In this paper we will further explore the geometrical properties of the three heavy-quark system.

Systems of heavy quarks are conveniently studied within an
effective field theory (EFT) framework, a treatment motivated by the
observation that these systems are non-relativistic and, therefore, 
characterized by, at least, three separated and hierarchically ordered energy scales: 
a hard scale of the order of the heavy-quark mass, $m$, a soft scale of the order of the typical 
relative momenta of the heavy quarks, which are much smaller than $m$, and an ultrasoft (US) scale 
of the order of the typical binding energy, which is much smaller than the relative momenta.\footnote{
In a three-body system, we may in general expect to have more than one typical relative momentum and more than one US energy scale.
To keep our discussion simple, we assume all relative momenta to be of the same order and so for all US energy scales. 
In the dynamical case, this is realized when the masses of the heavy quarks are of the same order.
In the static limit, which will be our main concern in the following, this condition 
is realized by locating the three quarks at distances of the same order. We emphasize that this condition 
may be (also largely) violated in different geometrical configurations.
}
We further assume that these scales are much larger than the typical hadronic scale $\Lambda_{\rm QCD}$, 
in this way justifying a perturbative treatment for all of them.
By integrating out modes associated with the different energy scales one goes through 
a sequence of EFTs~\cite{Brambilla:2004jw}: non-relativistic QCD (NRQCD), obtained from 
integrating out hard modes~\cite{Caswell:1985ui,Bodwin:1994jh} and potential non-relativistic QCD (pNRQCD), 
derived from integrating out gluons with soft momenta from NRQCD~\cite{Pineda:1997bj,Brambilla:1999xf}. 
Potential NRQCD provides a formulation of the non-relativistic system in terms of potentials 
and US interactions; for this reason it has proven a convenient framework for calculating US corrections.
Although originally designed for the study of $Q\bar Q$ bound states, i.e. quarkonia, 
pNRQCD has been subsequently applied also to baryons with two and three heavy quarks~\cite{Brambilla:2005yk,Brambilla:2009cd}.

In this paper we study the symmetry properties of three heavy-quark systems under exchange 
of the heavy-quark fields and their implications for the form of the pNRQCD Lagrangian.
We also calculate the US corrections of order $\alpha_{\rm s}^4\ln\alpha_{\rm s}$ 
to the singlet static energy and of order $\alpha_{\rm s}^4\ln\mu$ to the singlet static potential of a triple heavy baryon.
Whereas this has been achieved for the case of $Q\bar Q$ systems more than ten years ago~\cite{Brambilla:1999qa}, 
the result for $QQQ$ systems will be new.

The paper is organized as follows. Section~\ref{sec:construction} is devoted 
to set up pNRQCD for systems made of three static quarks. 
The explicit construction and color structure of the heavy-quark composite fields,
pNRQCD is conventionally formulated in, is outlined in detail. 
In Sec.~\ref{sec:sym}, we discuss the symmetry under exchange of the heavy-quark fields 
and analyze its implications for the various matching coefficients, i.e. the potentials, of pNRQCD.
In Sec.~\ref{sec:singlet}, we determine the correction of order $\alpha_{\rm s}^4\ln\alpha_{\rm s}$
to the singlet static energy. Restricting ourselves to an equilateral configuration of the heavy quarks, 
we finally solve in Sec.~\ref{sec:towardsmuinV} the renormalization group equations for the singlet, 
octet and decuplet static potentials at leading logarithmic accuracy. We conclude in Sec.~\ref{sec:conclusions}.

\section{$\text{p}$NRQCD for $QQQ$}
\label{sec:construction}
In this section, we shortly review the basic steps that lead to pNRQCD for systems made of three static quarks.
Finite mass corrections may be systematically added to the static Lagrangian in the form of irrelevant operators, 
some of which have been considered in~\cite{Brambilla:2005yk}. 
The non-relativistic nature of the system ensures that, apart from the kinetic energy, 
which is of the same order as the static potential, $1/m$ corrections are small.

\subsection{NRQCD}\label{subseq:NRQCD}
Our starting point is NRQCD in the static limit. In the quark sector the Lagrangian is identical 
with the heavy-quark effective theory Lagrangian~\cite{Eichten:1989zv} and reads
\begin{equation}
{\cal L}_{\rm NRQCD}=Q^{\dag}iD^0Q + \sum_{l}\bar{q}^{\,l}i\slashed{D}q^l - \frac{1}{4}F^a_{\mu\nu}F^{a\mu\nu} \,.
\label{eq:NRQCD}
\end{equation}
The heavy-quark fields $Q$ ($Q^{\dag}$), which annihilate (create) a heavy quark,
are described by Pauli spinors, whereas $q^l$ are the Dirac spinors that describe light (massless) quarks of flavor~$l$. 
The quantity $iD^0=i\partial^0-gA^0$ denotes the time component of the covariant derivative, where $g$ is
the strong gauge coupling, $\alpha_{\rm s} \equiv g^2/(4\pi)$, and $A^0$ is the time component of the gauge field.
The Lagrangian \eqref{eq:NRQCD} is insensitive to the flavor assignment of the heavy-quark fields,
a property known as heavy-quark symmetry.
We have omitted the heavy-antiquark sector, as it is irrelevant to our scope.

\subsection{pNRQCD}
For the purpose of studying heavy-quark bound states, it is 
convenient to employ an EFT where the heavy-quark potentials are explicit 
rather than encoded in dynamical gluons, as it is the case in NRQCD.
Such an EFT is pNRQCD, which is obtained from NRQCD by integrating out gluons 
whose momenta are soft. The degrees of freedom of pNRQCD are heavy-quark 
fields, light quarks and US gluons. As it is unnecessary to resolve the individual 
heavy quarks, pNRQCD is often formulated in terms of heavy-quark composite fields.  
The matching coefficients of pNRQCD multiplying operators bilinear in the composite fields 
may then be interpreted as the heavy-quark potentials in the corresponding color configurations.

The derivation of pNRQCD involves identifying the heavy-quark composite fields in NRQCD, 
matching them to pNRQCD, and explicitly ensuring that the resulting pNRQCD field content is ultrasoft. 
We start with the construction of the heavy-quark composite fields.
This is the point where the specific heavy-quark state that the EFT is meant to describe has to be specified. 
In our case, this is a $QQQ$ state.

\subsection{Geometry of a three-quark state}
\label{subseq:geomQQQ}
\begin{figure}[ht]
\begin{center}
 \includegraphics[width=4cm]{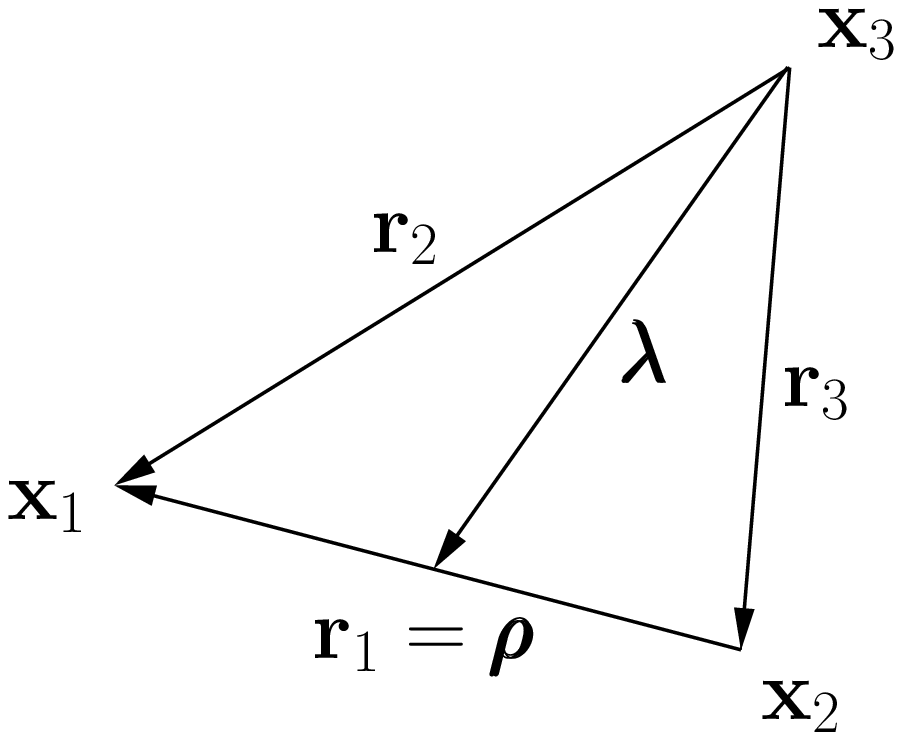}
\end{center}
\vspace*{-4mm}
\caption{Triangle formed by three heavy quarks 
located at the positions ${\bf x}_1$, ${\bf x}_2$ and ${\bf x}_3$. 
The vector $\pmb{\lambda}$ points from the heavy quark at ${\bf x}_3$ to 
the center of mass of the two heavy quarks at ${\bf x}_1$ and ${\bf x}_2$.}
\label{triangle}	
\end{figure}

To characterize the geometry of a $QQQ$ state, we call ${\bf x}_1$, ${\bf x}_2$ and ${\bf x}_3$ the 
positions of the quarks and define the vectors ${\bf r}_i$ ($i=1,2,3$) as follows (cf. Fig.~\ref{triangle}),
\begin{equation}
{\bf r}_1={\bf x}_1-{\bf x}_2\,, \qquad
{\bf r}_2={\bf x}_1-{\bf x}_3\,, \qquad
{\bf r}_3={\bf x}_2-{\bf x}_3\,.
\label{rx123}
\end{equation}
Note that the three vectors are not independent, for ${\bf r}_1+{\bf r}_3={\bf r}_2$.
Moreover, for three quarks of equal mass or static, it is useful to define the vectors 
\begin{equation}
\pmb{\rho}={\bf r}_1\,,\qquad
\pmb{\lambda}=\frac{{\bf r}_2+{\bf r}_3}{2}\,.
\label{rholambda}
\end{equation}

\subsection{Heavy-quark composite fields}\label{subseq:comp_fields}
Quarks transform under the fundamental representation, $3$, of the (color) gauge group SU(3)$_c$. 
Hence, a generic three (heavy) quark field made of fields located at the same point, 
$Q_iQ_jQ_k$ ($i,j,k=1,2,3$ denote color indices), transforms as a representation of $3\otimes3\otimes3$.  
The direct product can be decomposed into a sum of irreducible representations of SU(3)$_c$, namely
\begin{equation}
 3\otimes3\otimes3=1\oplus8\oplus8\oplus10\,. \label{prodinirreps}
\end{equation}
In general, however, the three quarks are located at different spatial positions ${\bf x}_1$, ${\bf x}_2$ and ${\bf x}_3$.
Under an SU(3)$_c$ gauge transformation, each heavy-quark field $ Q_i({\bf x},t)$ transforms as 
$\displaystyle Q_i({\bf x},t) \to U_{ii'}({\bf x},t)Q_{i'}({\bf x},t)$,
where $U({\bf x},t)=\exp\left[i\theta^a({\bf x},t)T^a\right]$,
and $T^a={\lambda^a}/{2}$ ($a=1, \ldots, 8$) denote the eight generators of SU(3)$_c$ in the fundamental representation; 
$\lambda^a$ are the Gell-Mann matrices.  
The decomposition (\ref{prodinirreps}) requires the fields to be linked to a common point~${\bf R}$.
For a multi-quark system a natural choice is the system's center of mass.
A way to link the quark fields to another point is through an equal-time straight Wilson string,
\begin{align}
\phi({\bf y},{\bf x},t)=
{\cal P}\exp\left\{ig\int_0^1{\rm d}s\ ({\bf y}-{\bf x})\cdot{\bf A}({\bf x}+({\bf y}-{\bf x})s,t)\right\},
\end{align}
where ${\bf A}={\bf A}^aT^a$ is the color gauge field, and ${\cal P}$ denotes path ordering of the color matrices.
Due to its transformation property under SU(3)$_c$ gauge transformations, 
$\phi({\bf y},{\bf x},t)$ $\to U({\bf y},t)\phi({\bf y},{\bf x},t)U^{\dag}({\bf x},t)$, 
the Wilson string acts as a gauge transporter, and 
$\phi({\bf R},{\bf x},t)Q({\bf x},t)$ $\to U({\bf R},t)\phi({\bf R},{\bf x},t)Q({\bf x},t)$
indeed transforms like a quark field located at ${\bf R}$. Hence, the following three-quark field,
\begin{align}
 \hspace*{-2.3mm}{\cal M}_{ijk}({\bf x}_1,{\bf x}_2,{\bf x}_3,t)=
\phi_{ii'}({\bf R},{\bf x}_1,t)Q_{i'}({\bf x}_1,t)\phi_{jj'}({\bf R},{\bf x}_2,t)Q_{j'}({\bf x}_2,t)
\phi_{kk'}({\bf R},{\bf x}_3,t)Q_{k'}({\bf x}_3,t), 
\label{QQQbzglR}
\end{align}
transforms as a $3\otimes3\otimes3$ representation of the SU(3)$_c$ gauge group, and, following
Eq.~(\ref{prodinirreps}), can be decomposed into a singlet, two octets
and a decuplet field with respect to gauge transformations in ${\bf R}$.

Since the quark fields do not commute, the order of the quark fields in Eq.~(\ref{QQQbzglR}) matters.
This observation will play a crucial role in Sec.~\ref{sec:sym}.
For simplicity, we have omitted an explicit reference to ${\bf R}$ in the argument of ${\cal M}$, 
which includes the time coordinate $t$ and the list of position coordinates 
(${\bf x}_1,{\bf x}_2,{\bf x}_3$) of the heavy-quark fields in the order (from left to right) 
of their appearance on the right-hand side of Eq.~(\ref{QQQbzglR}). 
The same convention is used for the color indices ($i,j,k$). 

The composite field ${\cal M}_{ijk}$ may be decomposed into a singlet, $S$, two octets, $O^{A}$ and  
$O^{S}$, and a decuplet, $\Delta$, according to
\begin{multline}
{\cal M}_{ijk}({\bf x}_1,{\bf x}_2,{\bf x}_3,t)=S({\bf x}_1,{\bf x}_2,{\bf x}_3,t){\underline {\bf S}}_{ijk}
+\sum_{a=1}^8O^{Aa}({\bf x}_1,{\bf x}_2,{\bf x}_3,t){\underline {\bf O}}^{Aa}_{ijk}  \\
+\sum_{a=1}^8O^{Sa}({\bf x}_1,{\bf x}_2,{\bf x}_3,t){\underline {\bf O}}^{Sa}_{ijk}
+\sum_{\delta=1}^{10}\Delta^{\delta}({\bf x}_1,{\bf x}_2,{\bf x}_3,t){\underline {\bf \Delta}}^{\delta}_{ijk}, 
\label{eq:dec}
\end{multline}
where ${\underline {\bf S}}_{ijk}$, ${\underline {\bf O}}^{Aa}_{ijk}$,  ${\underline {\bf O}}^{Sa}_{ijk}$ and 
${\underline {\bf \Delta}}^{\delta}_{ijk}$ are orthogonal and normalized  color tensors that satisfy the relations 
\begin{align}
&{\underline {\bf S}}_{ijk}{\underline {\bf S}}_{ijk}=1\,, \quad {\underline {\bf O}}^{Aa*}_{ijk}{\underline {\bf O}}^{Ab}_{ijk}
=\delta^{ab}\,, \quad {\underline {\bf O}}^{Sa*}_{ijk}{\underline {\bf O}}^{Sb}_{ijk}
=\delta^{ab}\,, \quad {\underline {\bf \Delta}}^{\delta}_{ijk}{\underline {\bf \Delta}}^{\delta'}_{ijk}=\delta^{\delta\delta'}\,, 
\nonumber\\
&{\underline {\bf S}}_{ijk}{\underline {\bf O}}^{Aa}_{ijk}={\underline {\bf S}}_{ijk}{\underline {\bf O}}^{Sa}_{ijk}
={\underline {\bf S}}_{ijk}{\underline {\bf \Delta}}^{\delta}_{ijk}
={\underline {\bf O}}^{Aa*}_{ijk}{\underline {\bf O}}^{Sb}_{ijk}
={\underline {\bf O}}^{Aa*}_{ijk}{\underline {\bf \Delta}}^{\delta}_{ijk}
={\underline {\bf O}}^{Sa*}_{ijk}{\underline {\bf \Delta}}^{\delta}_{ijk}=0\,, \label{OrthoN}
\end{align}
with $a,b\in \{1, \ldots, 8\}$, and $\delta,\delta'\in \{1, \ldots, 10\}$~\cite{Brambilla:2005yk}.
If the octet tensors ${\underline {\bf O}}^{Aa}_{ijk}$ and ${\underline {\bf O}}^{Sa}_{ijk}$ have the above properties, 
also the following linear combinations do,
\begin{align}
{\underline {\bf O}}'^{Aa}_{ijk}&={\rm e}^{{i}\varphi_A}\bigl({\underline {\bf O}}^{Aa}_{ijk}\cos\omega-{\underline {\bf O}}^{Sa}_{ijk}\sin\omega\bigr)\,, \\
{\underline {\bf O}}'^{Sa}_{ijk}&={\rm e}^{{i}\varphi_S}\bigl({\underline {\bf O}}^{Aa}_{ijk}\sin\omega+{\underline {\bf O}}^{Sa}_{ijk}\cos\omega\bigr)\,,
\end{align}
where $\omega$ is an arbitrary angle and $\varphi_A$, $\varphi_S$ denote generic phases. 
The octet tensors ${\underline {\bf O}}'^{Aa}_{ijk}$ and ${\underline {\bf O}}'^{Sa}_{ijk}$ 
hence form an alternative basis for the $8\oplus8$ sector. Requiring
\begin{align}
 O^{Aa}{\underline {\bf O}}^{Aa}_{ijk} + O^{Sa}{\underline {\bf O}}^{Sa}_{ijk}
= O'^{Aa} {\underline {\bf O}}'^{Aa}_{ijk} + O'^{Sa} {\underline {\bf O}}'^{Sa}_{ijk} \,,
\end{align}
the associated octet fields are related to the original ones through the dual relations
\begin{align}
 O'^{Aa}({\bf x}_1,{\bf x}_2,{\bf x}_3,t)&
= {\rm e}^{-{i}\varphi_A}\bigl[O^{Aa}({\bf x}_1,{\bf x}_2,{\bf x}_3,t)\cos\omega-O^{Sa}({\bf x}_1,{\bf x}_2,{\bf x}_3,t)\sin\omega\bigr]\,, \label{O_1} \\
 O'^{Sa}({\bf x}_1,{\bf x}_2,{\bf x}_3,t)&
= {\rm e}^{-{i}\varphi_S}\bigl[O^{Aa}({\bf x}_1,{\bf x}_2,{\bf x}_3,t)\sin\omega+O^{Sa}({\bf x}_1,{\bf x}_2,{\bf x}_3,t)\cos\omega\bigr]\,. \label{O_2}
\end{align}

To work out the pNRQCD Lagrangian explicitly, we choose a specific (matrix) representation of the color tensors, 
namely that given in~\cite{Brambilla:2005yk}, appendix B2. In order to keep this paper
self-contained, we reproduce it here. Sticking to this particular choice, the color-octet tensors are given by
\begin{align}
{\underline {\bf O}}^{Aa}_{ijk}=\frac{1}{2}\epsilon_{ijl}\lambda^a_{kl}\,, \label{O1}
\end{align}
and
\begin{align}
{\underline {\bf O}}^{Sa}_{ijk}=\frac{1}{2\sqrt{3}}\left(\epsilon_{jkl}\lambda^a_{il}+\epsilon_{ikl}\lambda^a_{jl}\right)\,. \label{O2}
\end{align}
The choice in Eqs.~(\ref{O1}) and (\ref{O2}) is such that ${\underline {\bf O}}^{Aa}_{ijk}$ 
and ${\underline {\bf O}}^{Sa}_{ijk}$ are antisymmetric and symmetric in the first two color indices,
respectively. Consequently, ${O}^{A}$ and ${O}^{S}$ will be referred to as the antisymmetric and symmetric octets.
Moreover, the color-singlet tensor ${\underline {\bf S}}_{ijk}$ is chosen to be totally antisymmetric,
\begin{align}
{\underline {\bf S}}_{ijk}=\frac{1}{\sqrt{6}}\epsilon_{ijk}\,, \label{S}
\end{align}
whereas the color-decuplet tensor ${\underline {\bf \Delta}}^{\delta}_{ijk}$ is totally symmetric (an alternative decuplet 
is in~\cite{Brambilla:2009cd}),
\begin{align}
{\underline {\bf \Delta}}^{1}_{111}&={\underline {\bf \Delta}}^{4}_{222}={\underline {\bf \Delta}}^{10}_{333}=1\,,
\quad\quad {\underline {\bf \Delta}}^{6}_{\{123\}}=\frac{1}{\sqrt{6}}\,, \nonumber\\
{\underline {\bf \Delta}}^{2}_{\{112\}}&={\underline {\bf \Delta}}^{3}_{\{122\}}={\underline {\bf \Delta}}^{5}_{\{113\}}
={\underline {\bf \Delta}}^{7}_{\{223\}}={\underline {\bf \Delta}}^{8}_{\{133\}}={\underline {\bf \Delta}}^{9}_{\{233\}}
=\frac{1}{\sqrt{3}}\,. \label{Delta}
\end{align}
The symbol $\{ijk\}$ denotes all permutations of the indices $ijk$; all components not listed explicitly in Eq.~(\ref{Delta}) 
are zero. Note that ${\underline {\bf S}}_{ijk}$ and ${\underline {\bf \Delta}}^{\delta}_{ijk}$ are real-valued quantities.

From Eq.~(\ref{QQQbzglR}), it follows that the three-quark field 
\begin{align}
 \Phi_{ijk}({\bf x}_1,{\bf x}_2,{\bf x}_3,t)\equiv Q_{i}({\bf x}_1,t)Q_{j}({\bf x}_2,t)Q_{k}({\bf x}_3,t)\,,  
\label{QQQselbst}
\end{align}
can be written as 
\begin{align}
 \Phi_{ijk}({\bf x}_1,{\bf x}_2,{\bf x}_3,t)
=\phi_{ii'}({\bf x}_1,{\bf R},t)\phi_{jj'}({\bf x}_2,{\bf R},t)\phi_{kk'}({\bf x}_3,{\bf R},t)
{\cal M}_{i'j'k'}({\bf x}_1,{\bf x}_2,{\bf x}_3,t)\,, 
\label{Phiijk}
\end{align}
where we have used that $\displaystyle \phi^{-1}({\bf y},{\bf x},t)=\phi^{\dag}({\bf y},{\bf x},t)=\phi({\bf x},{\bf y},t)$.
Finally, plugging Eq.~(\ref{eq:dec}) into Eq.~(\ref{Phiijk}) we may express the three-quark 
field $\Phi_{ijk}$ in terms of the composite singlet, octet and decuplet fields.  
The next step will consist in matching these composite fields to the corresponding ones in pNRQCD.

\subsection{Matching and multipole expansion}\label{subseq:match}
We denote with $|\Omega\rangle$ a generic Fock state containing no heavy quarks, but an
arbitrary number of US gluons and light quarks: $Q_i({\bf x},t)|\Omega\rangle=0$.
Therewith we define the three heavy-quark Fock state
\begin{equation}
|QQQ\rangle=\frac{1}{\cal N}\int{\rm d}^3x_1\int{\rm d}^3x_2\int{\rm d}^3x_3\;
\Phi_{ijk}({\bf x}_1,{\bf x}_2,{\bf x}_3,t)Q_k^{\dag}({\bf x}_3,t)Q_j^{\dag}({\bf x}_2,t)Q_i^{\dag}({\bf x}_1,t)
|\Omega\rangle\,, 
\label{proj}
\end{equation}
where ${\cal N}$ is a normalization factor and the composite field is now interpreted as independent of the heavy-quark fields. 
One can match NRQCD to pNRQCD by equating the expectation value 
of the NRQCD Hamiltonian in the state $|QQQ\rangle$ with the pNRQCD Hamiltonian
(see \cite{Pineda:1997bj,Brambilla:2004jw} for the matching in the $Q\bar{Q}$ case).
Thus, the heavy-quark fields in pNRQCD are cast into singlet, $S$, octet, $O^{Aa}$ and $O^{Sa}$, 
and decuplet, $\Delta^{\delta}$, fields. The gluons in pNRQCD are explicitly rendered US  
by multipole expanding the gluon fields in the relative coordinates ${\bf r}_i$ ($i=1,2,3$) 
with respect to the center of mass coordinate ${\bf R}$. The reason is that the center 
of mass coordinate (the ``location'' of the three heavy-quark system) scales like 
the inverse of the recoiling total momentum of the three quarks, which is of the order of the US energy scale, 
while the relative coordinates of the three quarks (describing the ``extension'' of the triple heavy baryon) 
scale like the inverse of the typical relative momenta of the heavy quarks, which are of the order of the soft scale.
As a result, ultrasoft gluons in pNRQCD are invariant under US gauge transformations, 
i.e. gauge transformations localized in ${\bf R}$.
A Legendre transform of the pNRQCD Hamiltonian finally provides us with the pNRQCD Lagrangian.

In the same way as NRQCD can be understood as an expansion of QCD 
in terms of the inverse of the heavy-quark masses, 
pNRQCD can be understood as an expansion of the gluon fields of NRQCD, 
projected onto the specific (two or three) heavy-quark Fock space, 
in powers of the relative coordinates of the heavy quarks.
Quantum corrections of the order of the soft scale are encoded in the 
matching coefficients of pNRQCD in the same way as quantum corrections of the 
order of the heavy-quark masses are encoded in the matching coefficients of NRQCD.
The matching coefficients of pNRQCD are typically non-analytic functions of the relative coordinates.

\subsection{The pNRQCD Lagrangian}
The pNRQCD Lagrangian is organized as an expansion in $1/m$ and in the relative coordinates ${\bf r}_i$.
Up to zeroth order in the $1/m$ expansion (static limit) and first order in the multipole expansion,
the pNRQCD Lagrangian for $QQQ$ systems reads 
\begin{align}
{\cal L}_{\rm pNRQCD}={\cal L}_{\rm pNRQCD}^{(0,0)}+{\cal L}_{\rm pNRQCD}^{(0,1)}\,. \label{LpNRQCD}
\end{align}
An explicit derivation of this Lagrangian can be found in \cite{Brambilla:2005yk}; here we recall its expression.
The term ${\cal L}_{\rm pNRQCD}^{(0,0)}$ describes at zeroth order in the multipole expansion 
the propagation of light quarks and US gluons as well as the temporal evolution of the static quarks, 
which are cast into singlet, $S\equiv S({\bf x}_1,{\bf x}_2,{\bf x}_3,t)$, octet, 
$O^A\equiv O^A({\bf x}_1,{\bf x}_2,{\bf x}_3,t)$ and $O^S\equiv O^S({\bf x}_1,{\bf x}_2,{\bf x}_3,t)$,   
and decuplet, $\Delta\equiv \Delta({\bf x}_1,{\bf x}_2,{\bf x}_3,t)$, fields (cf. Sec.~\ref{subseq:comp_fields}),
\begin{eqnarray}
{\cal L}_{\rm pNRQCD}^{(0,0)}&=&\int{\rm d}^3\!\rho\,{\rm d}^3\!\lambda\;\Bigl\{
S^{\dag}\left[i\partial_0-V^s\right]S+\Delta^{\dag}\left[iD_0-V^{d}\right]\Delta+O^{A\dag}\left[iD_0-V^o_A\right]O^A 
\nonumber\\
&&\hspace*{1.7cm}+O^{S\dag}\left[iD_0-V^o_S\right]O^S +O^{A\dag}\left[-V_{AS}^o\right]O^S+O^{S\dag}\left[-V_{AS}^o\right]O^A\Bigr\} 
\nonumber\\
&&+\sum_{l}\bar{q}^{\,l}i\slashed{D}q^l-\frac{1}{4}F^a_{\mu\nu}F^{a\mu\nu}\,.
\label{LpNRQCD1}
\end{eqnarray}
The matching coefficients $V^s$, $V^o_A$, $V^o_S$ and $V^d$ 
correspond to singlet, (antisymmetric and symmetric) octet and decuplet potentials. 
The coefficient $V^o_{AS}$ is an octet mixing potential.
The term ${\cal L}_{\rm pNRQCD}^{(0,1)}$ accounts for the interactions between static quarks and US gluons 
at first order in the multipole expansion, 
\begin{eqnarray}
{\cal L}_{\rm pNRQCD}^{(0,1)}&=&\int{\rm d}^3\!\rho\,{\rm d}^3\!\lambda\;\Bigl\{ V^{(0,1)}_{S\pmb{\rho}\cdot{\bf E}O^S}
\sum_{a=1}^8\tfrac{1}{2\sqrt{2}}\left[S^{\dag}\pmb{\rho}\cdot g{\bf E}^aO^{Sa}+O^{Sa\dag}\pmb{\rho}\cdot g{\bf E}^aS\right] 
\nonumber\\
&&\hspace*{1.75cm}-V^{(0,1)}_{S\pmb{\lambda}\cdot{\bf E}O^A}
\sum_{a=1}^8\tfrac{1}{\sqrt{6}}\left[S^{\dag}\pmb{\lambda}\cdot g{\bf E}^aO^{Aa}+O^{Aa\dag}\pmb{\lambda}\cdot g{\bf E}^aS\right] \nonumber\\
&&\hspace*{1.75cm}-V^{(0,1)}_{O^A\pmb{\lambda}\cdot{\bf E}O^A}
\sum_{a,b,c=1}^8\left(i\tfrac{f^{abc}}{6}+\tfrac{d^{\,abc}}{2}\right)O^{Aa\dag}\pmb{\lambda}\cdot g{\bf E}^bO^{Ac} \nonumber\\
&&\hspace*{1.75cm}+V^{(0,1)}_{O^S\pmb{\lambda}\cdot{\bf E}O^S}
\sum_{a,b,c=1}^8\left(i\tfrac{f^{abc}}{6}+\tfrac{d^{\,abc}}{2}\right)
O^{Sa\dag}\pmb{\lambda}\cdot g{\bf E}^bO^{Sc} \nonumber\\
&&\hspace*{1.75cm}-V^{(0,1)}_{O^A\pmb{\rho}\cdot{\bf E}O^S}
\sum_{a,b,c=1}^8\left(\tfrac{if^{abc}+3d^{\,abc}}{4\sqrt{3}}\right)
\left[O^{Aa\dag}\pmb{\rho}\cdot g{\bf E}^bO^{Sc}+O^{Sa\dag}\pmb{\rho}\cdot g{\bf E}^bO^{Ac}\right] \nonumber\\
&&\hspace*{1.75cm}+V^{(0,1)}_{O^A\pmb{\rho}\cdot{\bf E}\Delta}
\sum_{a,b=1}^8\sum_{\delta=1}^{10}\left[\left(
\epsilon_{ijk}T^a_{ii'}T^b_{jj'}
\underline{\pmb{\Delta}}^{\delta}_{i'j'k}\right)O^{Aa\dag}\pmb{\rho}\cdot g{\bf E}^b\Delta^{\delta}\right. \nonumber\\
&&\hspace*{4.8cm}\left.-\left(
\underline{\pmb{\Delta}}^{\delta}_{ijk}T^b_{ii'}T^a_{jj'}\epsilon_{i'j'k}\right)
\Delta^{\delta\dag}\pmb{\rho}\cdot g{\bf E}^bO^{Aa}\right] \nonumber\\
&&\hspace*{1.75cm}+V^{(0,1)}_{O^S\pmb{\lambda}\cdot{\bf E}\Delta}\sum_{a,b=1}^8\sum_{\delta=1}^{10}
\tfrac{2}{\sqrt{3}}\left[\left(
\epsilon_{ijk}T^a_{ii'}T^b_{jj'}
\underline{\pmb{\Delta}}^{\delta}_{i'j'k}\right)O^{Sa\dag}\pmb{\lambda}\cdot g{\bf E}^b\Delta^{\delta}\right. \nonumber\\
&&\hspace*{5.35cm}\left.-\left(
\underline{\pmb{\Delta}}^{\delta}_{ijk}T^b_{ii'}T^a_{jj'}\epsilon_{i'j'k}\right)\Delta^{\delta\dag}\pmb{\lambda}\cdot g{\bf E}^bO^{Sa}\right]\Bigr\},
\label{LpNRQCDusw}
\end{eqnarray}
where ${\bf E}={\bf E}^aT^a$ denotes the chromoelectric field evaluated at {\bf R} 
and the coefficients $V^{(0,1)}_{...}$ are matching coefficients associated to chromoelectric dipole interactions 
between $QQQ$ fields in different color representations.
The covariant derivatives, whose time components act on the octet and 
decuplet fields in Eq.~(\ref{LpNRQCD1}), are understood to
be in the octet and decuplet representations, respectively. 
They are given explicitly in appendix~\ref{app1}.

A mixing term, $-V_{AS}^o(O^{A\dag}O^S+O^{S\dag}O^A)$, has been included in ${\cal L}_{\rm pNRQCD}^{(0,0)}$.
Such a term was not considered in~\cite{Brambilla:2005yk}, but was first recognized in~\cite{Brambilla:2009cd}. 
The mixing potential will play a crucial role in the study of the symmetry of pNRQCD
under exchange of the heavy-quark fields (see Sec.~\ref{sec:sym}) and in the calculation of the US corrections 
to the singlet static energy (see Sec.~\ref{sec:singlet}).

For completeness, we list here the leading-order (LO) expressions for the various matching coefficients  
appearing in Eqs.~(\ref{LpNRQCD1}) and (\ref{LpNRQCDusw}). 
At order $\alpha_{\rm s}$
the potentials in Eq.~(\ref{LpNRQCD1}) are given by (cf.~\cite{Brambilla:2005yk},~\cite{Brambilla:2009cd})
\begin{eqnarray}
V^s({\bf r}_1,{\bf r}_2,{\bf r}_3)&=&
-\frac{2}{3}\alpha_{\rm s}\left(\frac{1}{|{\bf r}_1|}+\frac{1}{|{\bf r}_2|}+\frac{1}{|{\bf r}_3|}\right), 
\label{Vs}\\
V^d({\bf r}_1,{\bf r}_2,{\bf r}_3)&=&
\frac{1}{3}\alpha_{\rm s}\left(\frac{1}{|{\bf r}_1|}+\frac{1}{|{\bf r}_2|}+\frac{1}{|{\bf r}_3|}\right), 
\\
V^o_A({\bf r}_1,{\bf r}_2,{\bf r}_3)&=&
\alpha_{\rm s}\left(-\frac{2}{3}\frac{1}{|{\bf r}_1|}+\frac{1}{12}\frac{1}{|{\bf r}_2|}+\frac{1}{12}\frac{1}{|{\bf r}_3|}\right),
\label{VOA}\\
V^o_S({\bf r}_1,{\bf r}_2,{\bf r}_3)&=&
\alpha_{\rm s}\left(\frac{1}{3}\frac{1}{|{\bf r}_1|}-\frac{5}{12}\frac{1}{|{\bf r}_2|}-\frac{5}{12}\frac{1}{|{\bf r}_3|}\right),
\label{VOS}\\
V^o_{AS}({\bf r}_1,{\bf r}_2,{\bf r}_3)&=&
-\frac{\sqrt{3}}{4}\alpha_{\rm s}\left(\frac{1}{|{\bf r}_2|}-\frac{1}{|{\bf r}_3|}\right), 
\label{VAS}
\end{eqnarray}
whereas all matching coefficients in Eq.~(\ref{LpNRQCDusw}) are equal to one at LO.
The expressions for $V^s$ up to next-to-next-to-leading order (NNLO), and for $V^d$, $V^o_A$, $V^o_S$ and $V^o_{AS}$  
up to next-to-leading order (NLO) can be found in~\cite{Brambilla:2009cd} (the expression for $V^s$ up to NNLO is also 
in appendix~\ref{app2}).

\section{Symmetry under exchange of the heavy-quark fields}
\label{sec:sym}
As outlined in detail in Sec.~\ref{subseq:comp_fields}, the heavy-quark fields in the pNRQCD Lagrangian are written in terms of composite fields, 
which are proportional to $Q_{i}({\bf x}_1,t)Q_{j}({\bf x}_2,t)Q_{k}({\bf x}_3,t)$.
However, as there is no preferred ordering, and the heavy-quark fields anticommute, different orderings of the heavy quarks 
lead to different composite fields. The orderings are however arbitrary 
and the pNRQCD Lagrangian should be invariant under different 
orderings of the heavy-quark fields. We call this invariance symmetry under 
exchange of the heavy-quark fields or, in short, exchange symmetry. 
A special case of exchange symmetry is the symmetry under permutation 
of the heavy-quark fields. A different ordering of the heavy-quark fields 
can be realized either {\it (a)} by relabeling the heavy-quark coordinates in the pNRQCD Lagrangian 
or {\it (b)} by anticommuting the heavy-quark fields in the composite fields.
Since the two procedures lead to the same Lagrangian, this constrains the form of the heavy-quark potentials. 
In fact, the invariance of the Lagrangian under $(a)$ is trivially realized due to the additional integrations 
over the quark locations ${\rm x}_1$, ${\rm x}_2$ and ${\rm x}_3$, and only $(b)$ results in nontrivial transformations.

{\it (a)} We may relabel the coordinates  ${\bf x}_i$ and the relative vectors ${\bf r}_i$ 
in the pNRQCD Lagrangian according to one of the following two possibilities (other relabelings 
follow from these)
\begin{align}
{\bf x}_1\leftrightarrow{\bf x}_2\,,{\bf x}_3:\quad
\begin{cases}
{\bf r}_1\to-{\bf r}_1 \\
{\bf r}_2\to {\bf r}_3 \\
{\bf r}_3\to {\bf r}_2
\end{cases} \label{rwird1} 
,\\
{\bf x}_1\leftrightarrow{\bf x}_3\,,{\bf x}_2:\quad
\begin{cases}
{\bf r}_1\to-{\bf r}_3 \\
{\bf r}_2\to-{\bf r}_2 \\
{\bf r}_3\to-{\bf r}_1 
\end{cases} \label{rwird2} 
.
\end{align}
The relabelings affect the pNRQCD potentials and the ordering of the quark fields in the composite fields of pNRQCD. 

{\it (b)} Because the heavy-quark fields $Q_{i}({\bf x})$ of NRQCD satisfy equal-time anticommutation relations, 
$\{Q_{i}({\bf x},t),$ $Q_{j}({\bf y},t)\}=0$, from Eq.~(\ref{QQQselbst}) it follows that
\begin{eqnarray}
&& \Phi_{ijk}({\bf x}_1,{\bf x}_2,{\bf x}_3,t)=-\Phi_{jik}({\bf x}_2,{\bf x}_1,{\bf x}_3,t)\,, 
\label{soises0}\\
&& \Phi_{ijk}({\bf x}_1,{\bf x}_2,{\bf x}_3,t)=-\Phi_{kji}({\bf x}_3,{\bf x}_2,{\bf x}_1,t)\,. 
\label{soises1}
\end{eqnarray}  
These identities hold also for ${\cal M}_{ijk}({\bf x}_1,{\bf x}_2,{\bf x}_3,t)$, which is 
related to  $\Phi_{ijk}({\bf x}_1,{\bf x}_2,{\bf x}_3,t)$ through Eq.~(\ref{Phiijk}):
\begin{eqnarray}
&& {\cal M}_{ijk}({\bf x}_1,{\bf x}_2,{\bf x}_3,t)=-{\cal M}_{jik}({\bf x}_2,{\bf x}_1,{\bf x}_3,t)\,, 
\label{Msoises0}\\
&& {\cal M}_{ijk}({\bf x}_1,{\bf x}_2,{\bf x}_3,t)=-{\cal M}_{kji}({\bf x}_3,{\bf x}_2,{\bf x}_1,t)\,.
\label{Msoises1}
\end{eqnarray}
In turn, the identities for ${\cal M}_{ijk}({\bf x}_1,{\bf x}_2,{\bf x}_3,t)$ enable us to derive 
corresponding identities for the singlet, octet and decuplet fields just by multiplying Eqs.~(\ref{Msoises0}) and (\ref{Msoises1}) 
with ${\underline {\bf S}}_{ijk}$, ${\underline {\bf \Delta}}^{\delta}_{ijk}$, ${\underline {\bf O}}^{Aa*}_{ijk}$, 
or ${\underline {\bf O}}^{Sa*}_{ijk}$, respectively, and summing over $i,j,k$:
\begin{align}
\begin{cases}
S({\bf x}_1,{\bf x}_2,{\bf x}_3,t)             \hspace{-3mm} &= S({\bf x}_2,{\bf x}_1,{\bf x}_3,t) \\
\Delta^{\delta}({\bf x}_1,{\bf x}_2,{\bf x}_3,t) \hspace{-3mm} &=  -\Delta^{\delta}({\bf x}_2,{\bf x}_1,{\bf x}_3,t) \\
O^{Aa}({\bf x}_1,{\bf x}_2,{\bf x}_3,t)         \hspace{-3mm} &=  O^{Aa}({\bf x}_2,{\bf x}_1,{\bf x}_3,t) \\
O^{Sa}({\bf x}_1,{\bf x}_2,{\bf x}_3,t)         \hspace{-3mm} &=  -O^{Sa}({\bf x}_2,{\bf x}_1,{\bf x}_3,t) 
\label{block1}
\end{cases}
,
\end{align}
and 
\begin{align}
\begin{cases}
S({\bf x}_1,{\bf x}_2,{\bf x}_3,t)             \hspace{-3mm} &=  S({\bf x}_3,{\bf x}_2,{\bf x}_1,t)\\
\Delta^{\delta}({\bf x}_1,{\bf x}_2,{\bf x}_3,t) \hspace{-3mm} &= -\Delta^{\delta}({\bf x}_3,{\bf x}_2,{\bf x}_1,t)\\
O^{Aa}({\bf x}_1,{\bf x}_2,{\bf x}_3,t)         \hspace{-3mm} &= 
-
\tfrac{1}{2}
O^{Aa}({\bf x}_3,{\bf x}_2,{\bf x}_1,t) + 
\tfrac{\sqrt{3}}{2}
O^{Sa}({\bf x}_3,{\bf x}_2,{\bf x}_1,t)\\
O^{Sa}({\bf x}_1,{\bf x}_2,{\bf x}_3,t)         \hspace{-3mm} &= 
\tfrac{\sqrt{3}}{2}
O^{Aa}({\bf x}_3,{\bf x}_2,{\bf x}_1,t) + 
\tfrac{1}{2}
O^{Sa}({\bf x}_3,{\bf x}_2,{\bf x}_1,t)
\label{block2}
\end{cases}
. 
\end{align}
At variance with the relabeling {\it (a)}, anticommuting the heavy-quarks in the composite fields
only indirectly affects the pNRQCD potentials. Note that the octet transformations in \eqref{block1} and \eqref{block2} 
may be interpreted as a special case of the transformations~\eqref{O_1} and \eqref{O_2}
for $\varphi_S=0$, $\varphi_A=\pi$ and $\omega=\pi/3$.

By relabeling {\it (a)} or by anticommuting the heavy-quark fields {\it (b)} we get two 
versions of the pNRQCD Lagrangian that must be the same.
This requires the pNRQCD potentials to change in a well defined manner 
under the transformations \eqref{rwird1} and \eqref{rwird2}. 
In particular, if we restrict ourselves to the potentials in Eq.~(\ref{LpNRQCD1}), 
the singlet and decuplet potentials remain invariant, whereas the octet potentials transform as
\begin{align}
\begin{cases}
V^o_A(-{\bf r}_1,{\bf r}_3,{\bf r}_2)     \hspace{-3mm} &=   V^o_A({\bf r}_1,{\bf r}_2,{\bf r}_3) \\
V^o_S(-{\bf r}_1,{\bf r}_3,{\bf r}_2)     \hspace{-3mm} &=  V^o_S({\bf r}_1,{\bf r}_2,{\bf r}_3) \\
V_{AS}^o(-{\bf r}_1,{\bf r}_3,{\bf r}_2)   \hspace{-3mm} &=  - V_{AS}^o({\bf r}_1,{\bf r}_2,{\bf r}_3)
\label{Vrel1}
\end{cases}
, 
\end{align}
and 
\begin{align}
\begin{cases}
V^o_A(-{\bf r}_3,-{\bf r}_2,-{\bf r}_1)     \hspace{-3mm} &=  
\tfrac{1}{4}
V^o_A({\bf r}_1,{\bf r}_2,{\bf r}_3) 
+
\tfrac{3}{4}
V^o_S({\bf r}_1,{\bf r}_2,{\bf r}_3) 
- \tfrac{\sqrt{3}}{2}
V_{AS}^o({\bf r}_1,{\bf r}_2,{\bf r}_3) 
\\
V^o_S(-{\bf r}_3,-{\bf r}_2,-{\bf r}_1)     \hspace{-3mm} &=  
\tfrac{3}{4}
V^o_A({\bf r}_1,{\bf r}_2,{\bf r}_3) 
+
\tfrac{1}{4}
V^o_S({\bf r}_1,{\bf r}_2,{\bf r}_3) 
+
\tfrac{\sqrt{3}}{2}
V_{AS}^o({\bf r}_1,{\bf r}_2,{\bf r}_3) 
\\
V_{AS}^o(-{\bf r}_3,-{\bf r}_2,-{\bf r}_1)   \hspace{-3mm} &=  
\tfrac{\sqrt{3}}{4}
\bigl[V^o_S({\bf r}_1,{\bf r}_2,{\bf r}_3) -V^o_A({\bf r}_1,{\bf r}_2,{\bf r}_3)\bigr]
+\tfrac{1}{2}
V_{AS}^o({\bf r}_1,{\bf r}_2,{\bf r}_3)
\label{Vrel2}
\end{cases}
, 
\end{align}
for transformations of type \eqref{rwird1} and \eqref{rwird2} respectively.
We emphasize that the above transformations are general and do not 
rely on any specific geometry of the three quarks. They also do not rely on any perturbative expansion.
Furthermore, they are valid also beyond the static limit for any order in $1/m$.\footnote{
Note however that a generalization to finite heavy-quark masses, $m_1$, $m_2$ and $m_3$, 
would also require some adjustment in Eqs.~\eqref{rwird1} and \eqref{rwird2}, as -- besides the heavy-quark locations -- 
also the masses would have to be exchanged, e.g. in Eq.~\eqref{rwird1}, $m_1 \leftrightarrow m_2$, etc.}
As a simple application of the above formulas, let us consider for instance 
the LO expression of $V^o_{AS}({\bf r}_1,{\bf r}_2,{\bf r}_3)$ 
given in Eq.~(\ref{VAS}). Under (\ref{Vrel2}) it transforms into
\begin{align}
 V^o_{AS}(-{\bf r}_3,-{\bf r}_2,-{\bf r}_1) = 
-\frac{\sqrt{3}}{4}\alpha_{\rm s}\left(\frac{1}{|{\bf r}_2|}-\frac{1}{|{\bf r}_1|}\right)\,,
\label{eq:sojaso}
\end{align}
which is the result expected from relabeling the coordinates according to Eq.~\eqref{rwird2}. Let us emphasize again that the
inclusion of the octet mixing potential $V^o_{AS}$ in Eq.~\eqref{LpNRQCD1} is essential for reproducing 
the correct transformation properties of the octet potentials.  

Finally, it is interesting to apply relations~(\ref{Vrel1}) and (\ref{Vrel2})
to the most simple case of an equilateral geometry. 
In such a geometry we have a single length scale 
$r=|{\bf r}_1|=|{\bf r}_2|=|{\bf r}_3|$ and a single angle 
$\hat{\bf r}_1\cdot \hat{\bf r}_2 =-\hat{\bf r}_1\cdot\hat{\bf r}_3
=\hat{\bf r}_2\cdot\hat{\bf r}_3=\cos({\pi}/{3})$. Whenever 
the potentials are invariant under the transformations \eqref{rwird1} and \eqref{rwird2}, which is surely the case for two-body interactions 
but may not hold at higher orders, from  Eq.~(\ref{Vrel1}) it follows that $V_{AS}^o = 0$ and from Eq.~(\ref{Vrel2}) that 
\begin{align}
V^o_A (r)= V^o_S(r) \equiv V^o(r) \,.
\label{eq:Osgleich}
\end{align}

\section{The $QQQ$ singlet static energy at ${\cal O}(\alpha_{\rm s}^4\ln\alpha_{\rm s})$}
\label{sec:singlet}
The potentials of pNRQCD depend in general on a factorization scale $\mu$ separating soft from 
US contributions,\footnote{ 
This dependence, which will be displayed explicitly in the following, has been dropped in Eqs.~\eqref{Vrel1} and~\eqref{Vrel2}.
}
whereas the singlet static energy $E^s$ is an observable and therewith independent of $\mu$.
As in the $Q\bar Q$ case \cite{Brambilla:1999qa}, the $QQQ$ singlet static potential $V^s$ 
is expected to become $\mu$ dependent at next-to-next-to-next-to leading order (NNNLO), i.e. at order 
$\alpha_{\rm s}^4$~\cite{Brambilla:2005yk}.
The difference between the singlet static energy and the singlet static potential is encoded in an ultrasoft contribution 
denoted $\delta^s_{\rm US}$, which starts contributing at order $\alpha_s^4$. It depends on $\mu$ in such a way that $E^s$, given by
\begin{equation}
 E^s({\bf r}_1,{\bf r}_2,{\bf r}_3)=V^s({\bf r}_1,{\bf r}_2,{\bf r}_3;\mu)+\delta^s_{\rm US}({\bf r}_1,{\bf r}_2,{\bf r}_3;\mu), 
\label{E0}
\end{equation}
is $\mu$ independent. The cancelation of the $\mu$ dependence of $V^s$ against $\delta_{\rm US}^s$ at NNNLO 
leaves in $E^s$ a remnant, which is a contribution of order $\alpha_{\rm s}^4\ln\alpha_{\rm s}$. 
This is the leading perturbative contribution to $E^s$ that is non-analytic in $\alpha_{\rm s}$. 
The most convenient way to calculate the $\alpha_{\rm s}^4\ln\mu$ term in $V^s$,  
and the $\alpha_{\rm s}^4\ln\alpha_{\rm s}$ term in $E^s$, is by looking at the leading divergence of~$\delta_{\rm US}^s$. 
This requires the one-loop calculation of the color-singlet self energy as 
opposed to the three-loop calculation necessary to extract the term $\alpha_{\rm s}^4\ln\mu$ directly from $V^s$. 
We will perform this calculation in the following section.

\subsection{Determination of $\delta_{\rm US}^s$}
\label{detUS}
We aim at calculating  $\delta_{\rm US}^s$ up to order $\alpha_s^4$.
For this purpose we need the singlet and octet propagators, and 
the octet mixing potential at leading order [cf.~Eq.~(\ref{LpNRQCD1})],

\begin{center}
\includegraphics[width=11.5cm]{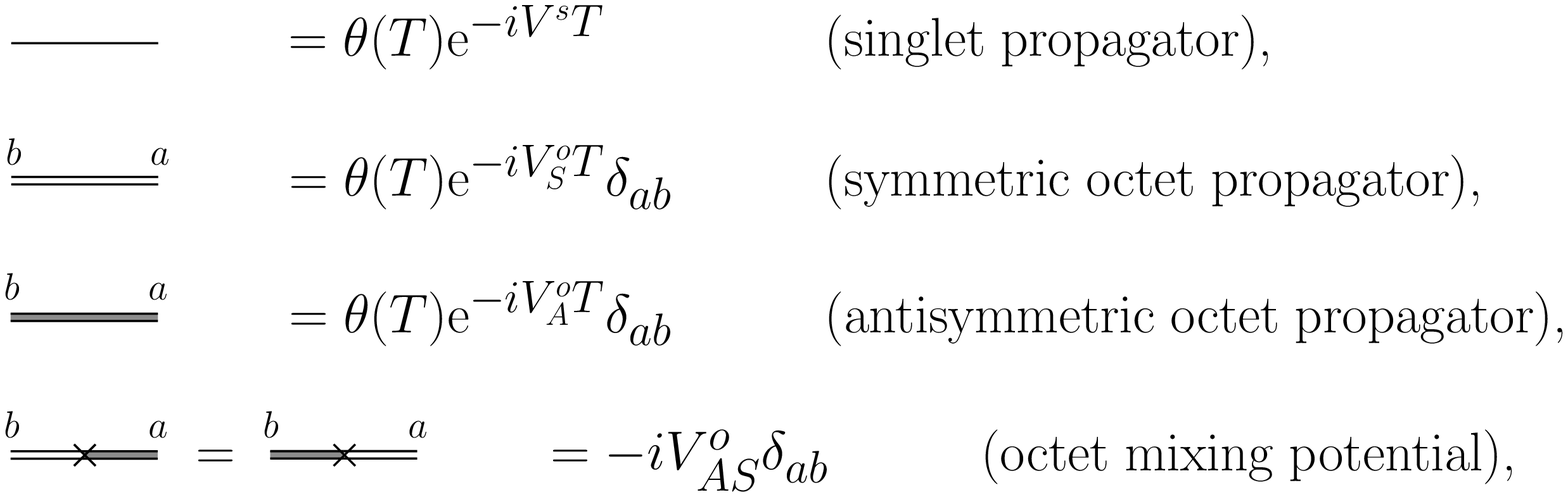}
\end{center}
\vspace*{-3cm}
\begin{minipage}{1\textwidth}
 \begin{align}
 \label{theprops} 
 \end{align}
\end{minipage}
\vspace*{1cm}\\

\noindent 
as well as the singlet-to-octet interaction vertices at order ${\bf r}_i$ 
in the multipole expansion [cf.~Eq.~(\ref{LpNRQCDusw}), note that the singlet 
couples differently to the symmetric and antisymmetric octets],

\begin{center}
\includegraphics[width=5.2cm]{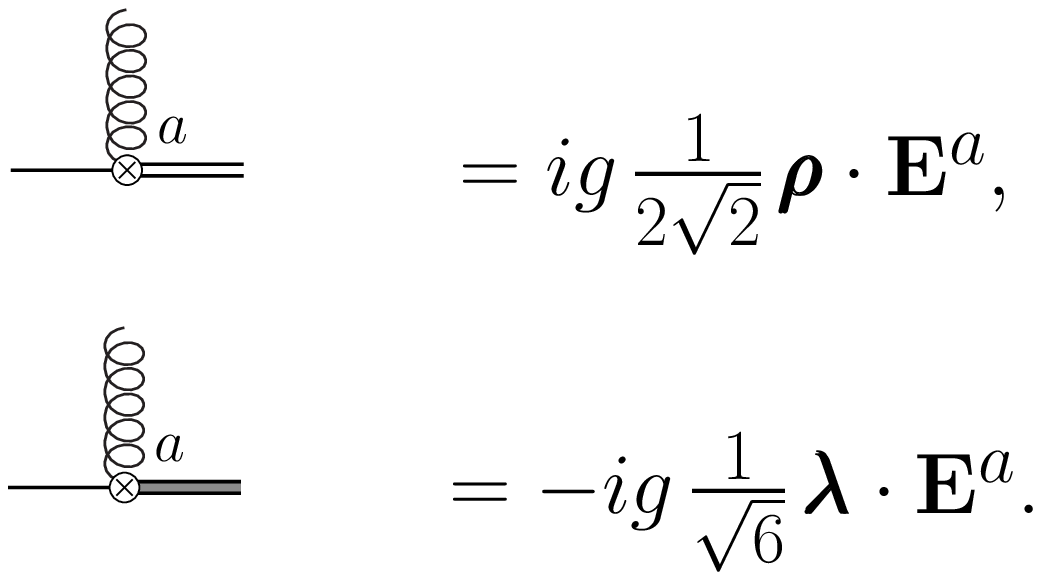}
\end{center}
\vspace*{-2.4cm}
\begin{minipage}{1\textwidth}
 \begin{align}
 \label{thevertices} 
 \end{align}
\end{minipage}
\vspace*{0.5cm}\\

\noindent 
The parameter $T$ in Eq.~(\ref{theprops}) is the propagation time.
The wavy lines in Eq.~(\ref{thevertices}) represent ultrasoft gluons;
note that we have written the vertices with US gluons treating the gluons as external fields.

The most noteworthy difference with respect to the calculation of $\delta_{\rm US}^s$
in the $Q\bar{Q}$ case is that here the singlet couples to two distinct 
octet fields and that the octet fields mix. For this reason the calculation in the baryonic 
case exhibits some novel features with respect to the analogous mesonic case.
Since the mixing of the octet fields is an effect of the same order as the energies of the octets, 
it must be accounted for to all orders when computing the physical octet-to-octet propagators.
The resummation of the octet mixing potential gives 
rise to three different types of resummed octet propagators:
\begin{itemize}
\item[(1)]{a resummed octet propagator, $G^o_{S}$, that describes the propagation from a 
symmetric initial state to a symmetric final state:

\begin{center}
\includegraphics[width=12cm]{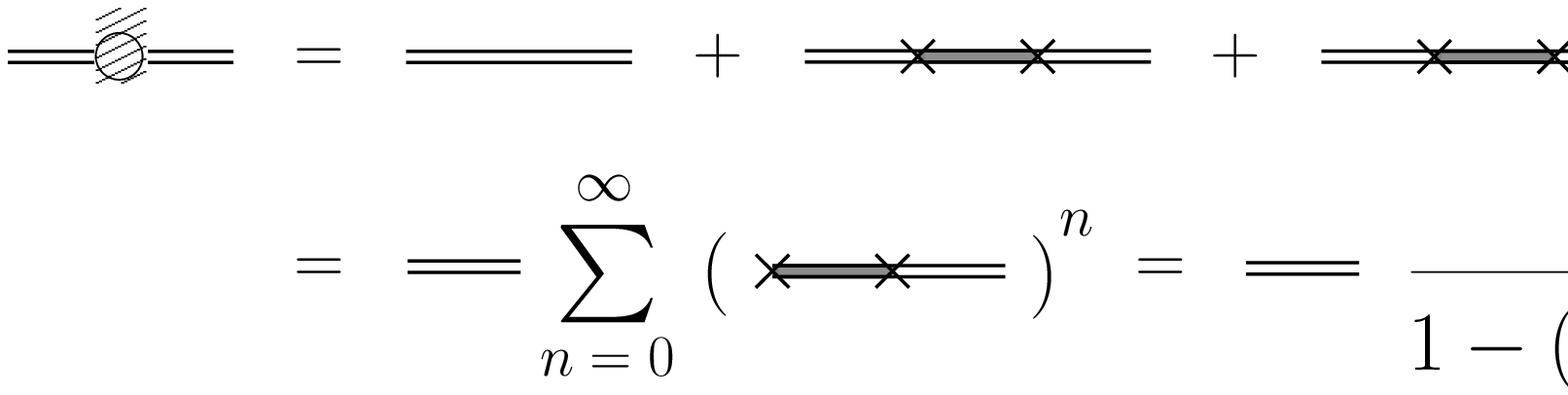}
\end{center}
}
\item[(2)]{a resummed octet propagator, $G^o_{A}$, that describes the propagation from an  
antisymmetric initial state to an antisymmetric final state:

\begin{center}
\includegraphics[width=11.1cm]{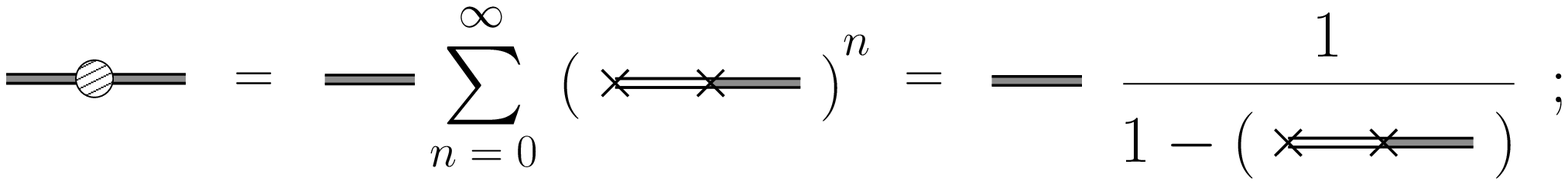}
\end{center}
}
\item[(3)]{a resummed octet propagator, $G^o_{AS}$, that describes the propagation from a 
symmetric initial state to an antisymmetric final state or vice versa:

\begin{center}
\includegraphics[width=5.3cm]{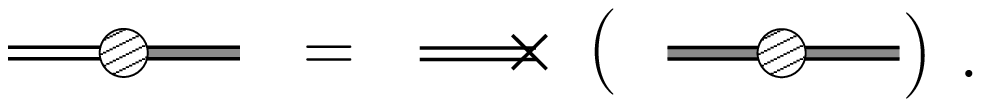}
\end{center}
}
\end{itemize}
The explicit expressions for the resummed octet propagators are most conveniently computed in momentum space 
and read 
\begin{align}
-i\left[G^o_{S}(E)\right]_{ab}&=\frac{i\delta_{ab}(E-V_A^o)}{(E-V_S^o+i\epsilon)(E-V_A^o+i\epsilon)-(V^{o}_{AS})^2}\,, \label{pe}\\
-i\left[G^o_{A}(E)\right]_{ab}&=\frac{i\delta_{ab}(E-V_S^o)}{(E-V_S^o+i\epsilon)(E-V_A^o+i\epsilon)-(V^{o}_{AS})^2}\,, \\
-i\left[G^o_{AS}(E)\right]_{ab}&=\frac{i\delta_{ab}V^{o}_{AS}}{(E-V_S^o+i\epsilon)(E-V_A^o+i\epsilon)-(V^{o}_{AS})^2}\,, \label{pa}
\end{align}
with $\epsilon\to0^+$. After performing a Fourier transform from energy $E$ to time $T$, we obtain

\begin{center}
\vspace*{3mm}
\includegraphics[width=12cm]{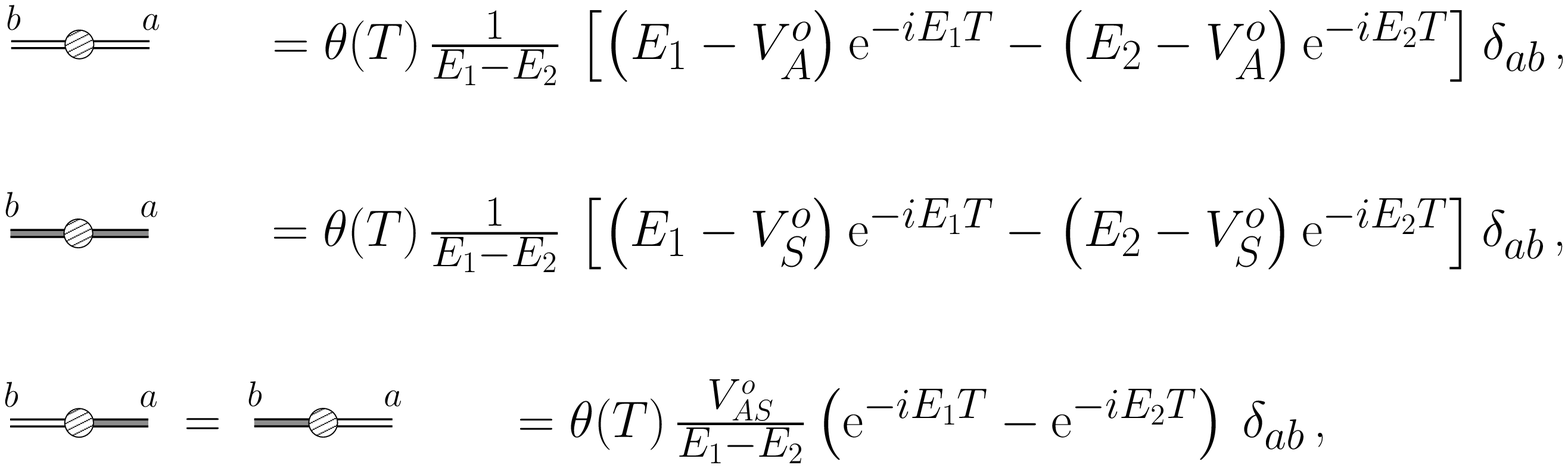}
\vspace*{-2mm}
\end{center}

\vspace*{-2.9cm}
\begin{minipage}{0.968\textwidth}
 \begin{align}
 \label{dressedprops} 
 \end{align}
\end{minipage}
\vspace*{1.9cm}

\pagebreak

where
\begin{equation}
E_{1,2}=\frac{V_A^o+V_S^o}{2}\pm\sqrt{\left(\frac{V_A^o-V_S^o}{2}\right)^2+(V^{o}_{AS})^2}-i\epsilon\,.
\label{E12}
\end{equation}
 
\begin{figure}[ht]
\begin{center}
\vspace*{5mm}
\includegraphics[width=10cm]{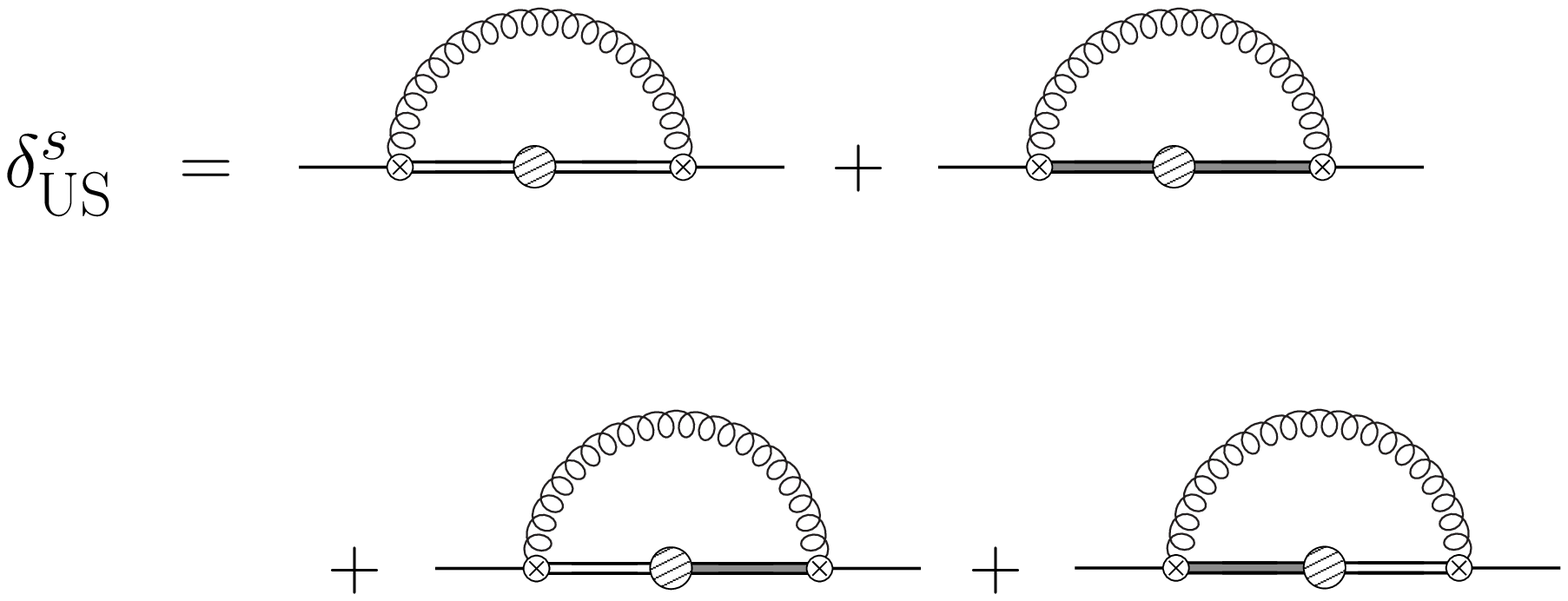}
\vspace*{-5mm}
\end{center}
\caption{Leading-order contributions to $\delta_{\rm US}^s$. As there is no direct
coupling between decuplet and singlet fields at first order in the multipole expansion, 
we do not have contributions involving decuplet degrees of freedom.}
\label{match}
\end{figure}

The US contribution $\delta_{\rm US}^s$ is given at LO by the color-singlet self-energy diagrams shown in Fig.~\ref{match}.
Because the singlet couples to two distinct octet fields and they mix, we have four such diagrams [cf. Eq.~\eqref{dressedprops}].
They give 
\begin{eqnarray}
\delta_{\rm US}^s &=&
- i g^2\left(\frac{1}{2\sqrt{2}}\right)^2\int_{0}^{\infty}{\rm d}t\, 
\frac{1}{E_1-E_2} \left[(E_1-V_A^o){\rm e}^{-it(E_1-V^s)}\right.
\nonumber \\
&&\hspace*{5.3cm}\left.-(E_2-V_A^o){\rm e}^{-it(E_2-V^s)}\right]\langle\pmb{\rho}\cdot{\bf E}^a(t)\pmb{\rho}\cdot{\bf E}^a(0)\rangle 
\nonumber \\
&& -i g^2\left(\frac{1}{\sqrt{6}}\right)^2\int_{0}^{\infty}{\rm d}t\,
\frac{1}{E_1-E_2}\left[(E_1-V_S^o){\rm e}^{-it(E_1-V^s)}\right. 
\nonumber \\
&&\hspace*{5.1cm}\left.-(E_2-V_S^o){\rm e}^{-it(E_2-V^s)}\right]\langle\pmb{\lambda}\cdot{\bf E}^a(t)\pmb{\lambda}\cdot{\bf E}^a(0)\rangle 
\nonumber \\
&&
+2ig^2\frac{1}{2\sqrt{2}}\frac{1}{\sqrt{6}}
\int_{0}^{\infty}{\rm d}t\,\frac{V^o_{AS}}{E_1-E_2}\left[{\rm e}^{-it(E_1-V^s)}\right. 
\left.-{\rm e}^{-it(E_2-V^s)}\right]
\langle\pmb{\rho}\cdot{\bf E}^a(t)\pmb{\lambda}\cdot{\bf E}^a(0)\rangle ,
\label{pNRQCDdiags}
\end{eqnarray}
where $\langle \cdots \rangle$ stands for a vacuum expectation value.
In writing the various contributions in Eq.~\eqref{pNRQCDdiags}, we have kept the same order as 
in Fig.~\ref{match}: the first two terms correspond to the two diagrams shown 
in the first line of Fig.~\ref{match}, and the last contribution is the sum of the two diagrams 
in the second line of Fig.~\ref{match}, which are equal.

The vacuum expectation value of two chromoelectric fields reads
in dimensional regularization ($d = 4-2\varepsilon$ is the number of dimensions)
\begin{equation}
\langle{\bf a}\cdot{\bf E}^a(t){\bf b}\cdot{\bf E}^a(0)\rangle=
{\bf a}\cdot{\bf b}\,\frac{4(d-2)}{(d-1)}\mu^{4-d}\int\frac{{\rm d}^{d-1}q}{(2\pi)^{d-1}}|{\bf q}|{\rm e}^{-i|{\bf q}|t}+{\cal O}(\alpha_s)\,,
\end{equation}
where ${\bf a}$ and ${\bf b}$ are two generic vectors and $t>0$.
Performing the integrals in (\ref{pNRQCDdiags}) we obtain 
\begin{eqnarray}
\delta_{\rm US}^s = \frac{4}{3}\frac{\alpha_{\rm s}}{\pi}\frac{1}{E_1-E_2}\hspace*{-3mm}
&& 
\left[\left(\frac{|\pmb{\rho}|^2}{4}(E_1-V_A^o)+\frac{|\pmb{\lambda}|^2}{3}(E_1-V_S^o)
-\frac{\pmb{\rho}\cdot\pmb{\lambda}}{\sqrt{3}}V^{o}_{AS}\right)(E_1-V^s)^3\right. 
\nonumber\\
&&\hspace*{1cm}
\times\left(\frac{1}{\varepsilon}-\gamma_E-\ln\frac{(E_1-V^s)^2}{\pi\mu^2}+\frac{5}{3}\right) 
\nonumber\\
&&
-\left(\frac{|\pmb{\rho}|^2}{4}(E_2-V_A^o)+\frac{|\pmb{\lambda}|^2}{3}(E_2-V_S^o)
-\frac{\pmb{\rho}\cdot\pmb{\lambda}}{\sqrt{3}}V^{o}_{AS}\right)(E_2-V^s)^3 \nonumber\\
&&
\hspace*{1cm}\left.
\times\left(\frac{1}{\varepsilon}-\gamma_E -\ln\frac{(E_2-V^s)^2}{\pi\mu^2}+\frac{5}{3}\right)\right]\,, 
\label{US1}
\end{eqnarray}
where $\gamma_E$ is the Euler--Mascheroni constant.
Equation~\eqref{US1} comprises the entire US contribution up to order $\alpha_s^4$. The explicit expressions may be 
obtained by replacing $E_1$ and $E_2$ with the right-hand side of Eq. (\ref{E12}), and $V^s$, $V_A^o$, $V_S^o$ and $V_{AS}^o$ 
by the LO expressions given in Eqs.~(\ref{Vs}), (\ref{VOA}), (\ref{VOS}) and (\ref{VAS}) respectively.  
Equation~(\ref{US1}) corrects the expression derived in~\cite{Brambilla:2005yk} where the mixing 
of the octet fields was not taken into account. Hence, the result of~\cite{Brambilla:2005yk} 
is retained from Eq.~\eqref{US1} by setting $V_{AS}^o=0$.

\subsection{Invariance of $\delta_{\rm US}^s$ under exchange symmetry}
\label{sec:invdeltaUS}
The US correction, $\delta_{\rm US}^s$, calculated in the previous section is expected to be invariant 
under the exchange symmetry discussed in Sec.~\ref{sec:sym}.
To verify this we observe that according to  Eqs.~(\ref{Vrel1}) and (\ref{Vrel2}) the combinations 
$(V_A^o+V_S^o)$ and $\left[\left(V_A^o-V_S^o\right)^2/4+(V^{o}_{AS})^2\right]$ are each invariant.
This implies that both $E_1$ and $E_2$ are invariant according to the definition (\ref{E12}).
Also the singlet static potential, $V^s$, is invariant at LO [see Eq.~(\ref{Vs})].
If we rewrite explicitly the expression ${|\pmb{\rho}|^2}/{4}+{|\pmb{\lambda}|^2}/{3}$ in terms 
of the positions of the heavy quarks with the help of Eqs.~(\ref{rx123}) and~(\ref{rholambda}),  
\begin{equation}
\frac{|\pmb{\rho}|^2}{4}+\frac{|\pmb{\lambda}|^2}{3}=
\frac{1}{3}\left({\bf x}_1^2+{\bf x}_2^2+{\bf x}_3^2-{\bf x}_1\cdot{\bf x}_2-{\bf x}_1\cdot{\bf x}_3-{\bf x}_2\cdot{\bf x}_3\right), 
\label{A}
\end{equation}
it is evident that this expression is invariant under the transformations \eqref{rwird1} and \eqref{rwird2}.
Finally, we have to show that the expression 
\begin{equation}
V^o_A \frac{|\pmb{\rho}|^2}{4} +  V^o_S\frac{|\pmb{\lambda}|^2}{3} 
+ V^o_{AS}\frac{\pmb{\rho}\cdot\pmb{\lambda}}{\sqrt{3}}, 
\label{B}
\end{equation}
is also invariant. This is a straightforward, although not manifest, consequence of the transformations 
(\ref{rwird1}), (\ref{rwird2}), (\ref{Vrel1}) and (\ref{Vrel2}), which completes the proof that $\delta_{\rm US}^s$ is invariant 
under the exchange symmetry. The invariance of $\delta_{\rm US}^s$ is directly inherited by the contribution 
to $V^s$ at order $\alpha_{\rm s}^4\ln \mu$ and the singlet static energy $E^s$ at order $\alpha_{\rm s}^4\ln \alpha_{\rm s}$.

\subsection{The $QQQ$ Singlet Static Potential and Energy}
According to Eq.~(\ref{E0}), the divergence and the $\alpha_{\rm s}^4\ln\mu$ term in $\delta_{\rm US}^s$ 
must cancel against a divergence and a term $\alpha_{\rm s}^4\ln\mu$ in the singlet static potential $V^s$.
Therefore the  $\alpha_{\rm s}^4\ln\mu$ part of the potential may be read off from Eq. (\ref{US1}). 
In a minimal subtraction scheme, the singlet static potential up to order $\alpha_{\rm s}^4\ln\mu$ is then given by 
\begin{eqnarray}
V^s({\bf r}_1,{\bf r}_2,{\bf r}_3;\mu) &=& V^s_{\rm NNLO}({\bf r}_1,{\bf r}_2,{\bf r}_3)
\nonumber\\
- \frac{\alpha_{\rm s}^4}{3\pi}\ln\mu && \hspace{-6mm}
\left[
\left({\bf r}_1^2+\frac{({\bf r}_2+{\bf r}_3)^2}{3}\right)
\left(\frac{1}{|{\bf r}_1|^2} + \frac{1}{|{\bf r}_2|^2} + \frac{1}{|{\bf r}_3|^2} 
-\frac{1}{4}\frac{|{\bf r}_1| + |{\bf r}_2| + |{\bf r}_3|}{|{\bf r}_1||{\bf r}_2||{\bf r}_3|}\right)
\right.
\nonumber\\
&&
\hspace{3.4cm}
\times \left(\frac{1}{|{\bf r}_1|} + \frac{1}{|{\bf r}_2|} + \frac{1}{|{\bf r}_3|} \right)
\nonumber\\
&&
\hspace{-6mm}
+
\left({\bf r}_1^2-\frac{({\bf r}_2+{\bf r}_3)^2}{3}\right)
\left(\frac{1}{|{\bf r}_1|^2} + \frac{1}{|{\bf r}_2|^2} + \frac{1}{|{\bf r}_3|^2} 
+\frac{5}{4}\frac{|{\bf r}_1| + |{\bf r}_2| + |{\bf r}_3|}{|{\bf r}_1||{\bf r}_2||{\bf r}_3|}\right)
\nonumber\\
&&
\hspace{3.4cm}
\times \left(\frac{1}{|{\bf r}_1|} - \frac{1}{2|{\bf r}_2|} - \frac{1}{2|{\bf r}_3|} \right)
\nonumber\\
&&
\hspace{-6mm}
+
{\bf r}_1\cdot({\bf r}_2+{\bf r}_3)
\left(\frac{1}{|{\bf r}_1|^2} + \frac{1}{|{\bf r}_2|^2} + \frac{1}{|{\bf r}_3|^2} 
+\frac{5}{4}\frac{|{\bf r}_1| + |{\bf r}_2| + |{\bf r}_3|}{|{\bf r}_1||{\bf r}_2||{\bf r}_3|}\right)
\nonumber\\
&&
\hspace{3.4cm}
\left.
\times \left(\frac{1}{|{\bf r}_2|} - \frac{1}{|{\bf r}_3|} \right)
\right]
\,.
\label{Vs3loop}
\end{eqnarray}
The singlet static potential up to order  $\alpha_{\rm s}^3$, which we have denoted by $V^s_{\rm NNLO}$, has been calculated in 
Ref.~\cite{Brambilla:2009cd} and is reproduced in appendix~\ref{app2}. At order $\alpha_{\rm s}^3$,  $V^s_{\rm NNLO}$ contains 
the leading three-body potential; also the new term proportional to $\alpha_{\rm s}^4\ln\mu$ 
that we have added here is a genuine three-body potential.

Summing up the singlet static potential (\ref{Vs3loop}) with the US contribution (\ref{US1}) we obtain 
the singlet static energy up to order $\alpha_{\rm s}^4\ln\alpha_{\rm s}$, which reads
\begin{eqnarray}
E^s({\bf r}_1,{\bf r}_2,{\bf r}_3) &=& V^s_{\rm NNLO}({\bf r}_1,{\bf r}_2,{\bf r}_3)
\nonumber\\
- \frac{\alpha_{\rm s}^4}{3\pi}\ln\alpha_{\rm s} && \hspace{-6mm}
\left[
\left({\bf r}_1^2+\frac{({\bf r}_2+{\bf r}_3)^2}{3}\right)
\left(\frac{1}{|{\bf r}_1|^2} + \frac{1}{|{\bf r}_2|^2} + \frac{1}{|{\bf r}_3|^2} 
-\frac{1}{4}\frac{|{\bf r}_1| + |{\bf r}_2| + |{\bf r}_3|}{|{\bf r}_1||{\bf r}_2||{\bf r}_3|}\right)
\right.
\nonumber\\
&&
\hspace{3.4cm}
\times \left(\frac{1}{|{\bf r}_1|} + \frac{1}{|{\bf r}_2|} + \frac{1}{|{\bf r}_3|} \right)
\nonumber\\
&&
\hspace{-6mm}
+
\left({\bf r}_1^2-\frac{({\bf r}_2+{\bf r}_3)^2}{3}\right)
\left(\frac{1}{|{\bf r}_1|^2} + \frac{1}{|{\bf r}_2|^2} + \frac{1}{|{\bf r}_3|^2} 
+\frac{5}{4}\frac{|{\bf r}_1| + |{\bf r}_2| + |{\bf r}_3|}{|{\bf r}_1||{\bf r}_2||{\bf r}_3|}\right)
\nonumber\\
&&
\hspace{3.4cm}
\times \left(\frac{1}{|{\bf r}_1|} - \frac{1}{2|{\bf r}_2|} - \frac{1}{2|{\bf r}_3|} \right)
\nonumber\\
&&
\hspace{-6mm}
+
{\bf r}_1\cdot({\bf r}_2+{\bf r}_3)
\left(\frac{1}{|{\bf r}_1|^2} + \frac{1}{|{\bf r}_2|^2} + \frac{1}{|{\bf r}_3|^2} 
+\frac{5}{4}\frac{|{\bf r}_1| + |{\bf r}_2| + |{\bf r}_3|}{|{\bf r}_1||{\bf r}_2||{\bf r}_3|}\right)
\nonumber\\
&&
\hspace{3.4cm}
\left.
\times \left(\frac{1}{|{\bf r}_2|} - \frac{1}{|{\bf r}_3|} \right)
\right]
\,.
\label{eq:E0full}
\end{eqnarray}
The logarithm of $\alpha_{\rm s}$ signals that an ultraviolet divergence from the US scale 
has canceled against an infrared divergence from the soft scale.

Finally, it may be useful to express Eqs.~\eqref{Vs3loop} and \eqref{eq:E0full} in a way that makes manifest 
the invariance under exchange symmetry proven in Sec.~\ref{sec:invdeltaUS}.
First, we recall that ${\bf r}_1$, ${\bf r}_2$ and ${\bf r}_3$ are not independent 
(cf. Sec.~\ref{subseq:geomQQQ}) and write
\begin{equation}
E^s({\bf r}_1,{\bf r}_2,{\bf r}_3)=E^s({\bf r}_2-{\bf r}_3,{\bf r}_2,{\bf r}_3)\equiv E^s({\bf r}_2,{\bf r}_3), 
\end{equation}
then we observe that 
\begin{equation}
E^s({\bf r}_2,{\bf r}_3)=E^s({\bf r}_3,{\bf r}_2).
\end{equation}
Hence an expression of the singlet static energy, which is manifestly invariant under exchange symmetry, is
\begin{equation}
E^s({\bf r}_1,{\bf r}_2,{\bf r}_3)=\frac{E^s({\bf r}_2,{\bf r}_3)+E^s({\bf r}_1,-{\bf r}_3)+E^s(-{\bf r}_2,-{\bf r}_1)}{3}.
\end{equation}
Similarly one can obtain a manifestly invariant expression of the singlet static potential.

\section{Renormalization group improvement of the singlet static potential in an equilateral geometry}
\label{sec:towardsmuinV}
The US logarithms that start appearing in the static potential at NNNLO may be resummed 
to all orders by solving the corresponding renormalization group equations.
These are a set of equations that describe the scale dependence of the static potentials in the different color representations.
They follow from requiring that the static energies of the $QQQ$ system and its gluonic excitations 
are independent of the renormalization scheme. The potentials in the different color representations mix under renormalization.
This may be easily understood by looking at the renormalization group equation for the 
singlet potential that can be derived from $\mu\,{\rm d}V^s/{\rm d}\mu = - \mu\,{\rm d}\delta_{\rm US}^s/{\rm d}\mu$ and Eq.~\eqref{US1},
\begin{eqnarray}
 \mu\frac{\rm d}{{\rm d}\mu}V^s&=&-\frac{8}{3}\frac{\alpha_{\rm s}}{\pi}
\left\{\left[\frac{V^o_S-V^o_A}{2}\left(\frac{|\pmb{\rho}|^2}{4}-\frac{|\pmb{\lambda}|^2}{3}\right)
-V^o_{AS}\,\frac{\pmb{\rho}\cdot\pmb{\lambda}}{\sqrt{3}}\right]\right. 
\nonumber\\
&& \hspace{3cm}
\times\left[3\,\left(\frac{V^o_S+V^o_A}{2}-V^s\right)^2+\frac{(V^o_S-V^o_A)^2}{4} + (V^o_{AS})^2\right]
\nonumber\\
&&
\hspace*{1.15cm}+\left(\frac{V^o_S+V^o_A}{2}-V^s\right)\left(\frac{|\pmb{\rho}|^2}{4} +\frac{|\pmb{\lambda}|^2}{3}\right)
\nonumber\\
&& \hspace{3cm}
\times\left.\left[\left(\frac{V^o_S+V^o_A}{2}-V^s\right)^2+3\frac{(V^o_S-V^o_A)^2}{4} + 3(V^o_{AS})^2\right]\right\}\,.
\label{eq:RgVsgen}
\end{eqnarray}
It shows the explicit dependence of the running of $V^s$ on the octet potentials and octet mixing potential.

In the $Q\bar{Q}$ case the renormalization group equations have been solved for the singlet 
static potential at next-to-next-to-leading logarithmic (NNLL) accuracy in~\cite{Pineda:2000gza} 
and at next-to-next-to-next-to-leading logarithmic (NNNLL) accuracy in~\cite{Brambilla:2009bi}.\footnote{
An NNLL accuracy amounts at resumming $\alpha_{\rm s}^3 (\alpha_{\rm s} \ln \mu)^n$ terms 
and an NNNLL accuracy amounts at resumming $\alpha_{\rm s}^4 (\alpha_{\rm s} \ln \mu)^n$ terms, with $n\in\mathbb{N}_0$.} 
In the $QQQ$ case similar results can be obtained by solving Eq.~\eqref{eq:RgVsgen} with the corresponding renormalization group 
equations for the octet and decuplet potentials. There is however a difference between the $Q\bar{Q}$ 
and the  $QQQ$ case that is worth highlighting. While in a $Q\bar{Q}$ system there is just one length, 
the distance between the heavy quark and antiquark, the generic three-body system is characterized by more than one length. 
For a general three-body geometry, therefore, logarithmic corrections in the US scale 
could be numerically as important as finite logarithms involving ratios among the different lengths of the system. 
The calculation of these finite logarithms requires the calculation of the $QQQ$ static Wilson loop. 
However, these logarithms are unimportant if the distances between the heavy quarks are similar. 
In the following, we will therefore restrict ourselves to the simplest case 
of three static quarks located at the corners of an equilateral triangle.
In this situation, the three-body system is characterized, like the two-body one, by just one fundamental length, 
which can be identified with the length of each side of the triangle: $|{\bf r}_1|=|{\bf r}_2|=|{\bf r}_3|=r$. 

\begin{figure}[ht]
\begin{center}
\vspace*{5mm}
\includegraphics[width=15cm]{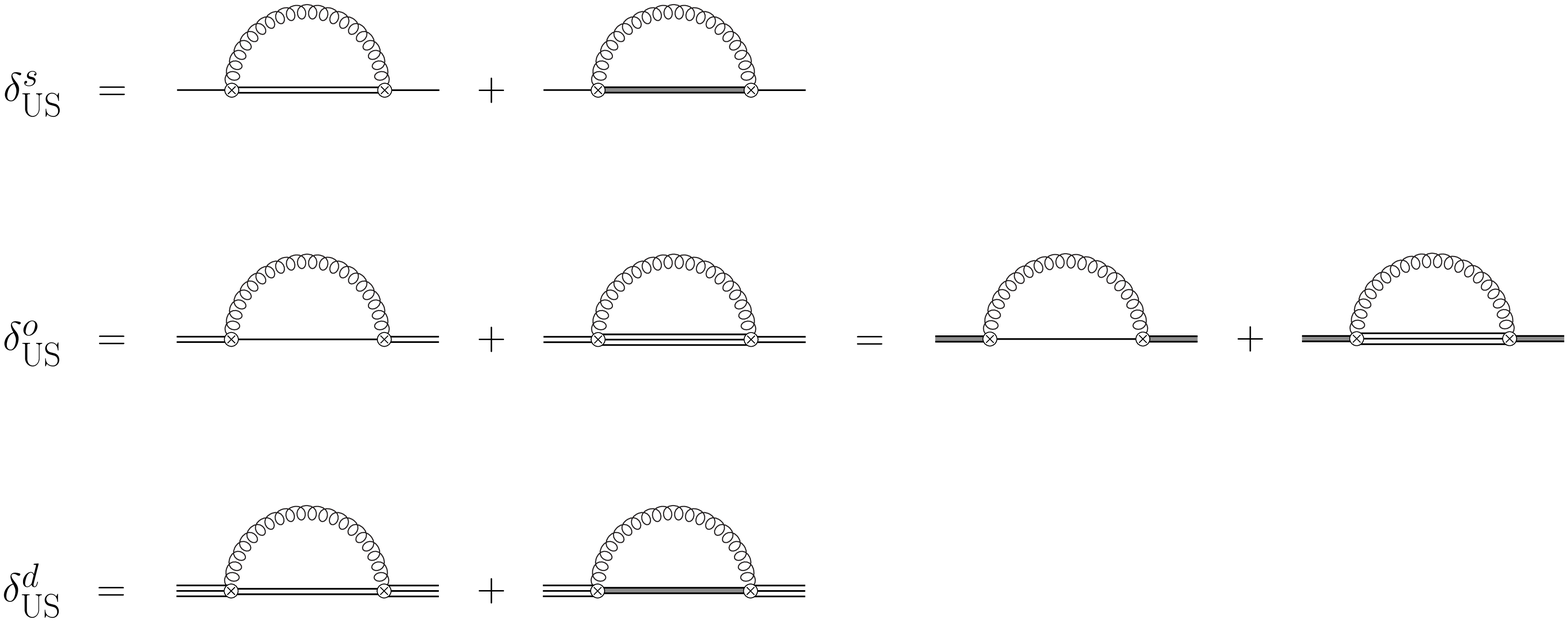}
\end{center}
\vspace*{-5mm}
\caption{
Leading-order ultrasoft contributions to the singlet, $\delta_{\rm US}^s$, octet, $\delta_{\rm US}^o$, and 
decuplet,  $\delta_{\rm US}^d$, energies in an equilateral geometry.
The triple lines represent the decuplet propagator, $\theta(T)e^{-iV^dT}\delta_{\delta\delta'}$; 
the decuplet can couple to a symmetric octet, with vertex 
$ig\tfrac{2}{\sqrt{3}}\left(\epsilon_{ijk}T^a_{ii'}T^b_{jj'}\underline{\pmb{\Delta}}^{\delta}_{i'j'k}\right)\pmb{\lambda}\cdot {\bf E}^b$, 
or to an antisymmetric octet, with vertex 
$ig\left(\epsilon_{ijk}T^a_{ii'}T^b_{jj'}\underline{\pmb{\Delta}}^{\delta}_{i'j'k}\right)\pmb{\rho}\cdot{\bf E}^b$; 
the other propagators and vertices have been introduced in Eqs.~(\ref{theprops}) and (\ref{thevertices}).}
\label{equimatch}	
\end{figure}

In the equilateral limit at least up to NLO, the different octet fields do not mix, moreover, 
as has been shown in Eq.~(\ref{eq:Osgleich}), the two octet potentials $V_S^o$ and $V_A^o$ are equal.
The US contribution for the singlet static energy follows by specializing the general formula \eqref{US1} to the equilateral limit. 
The US contributions for the octet and decuplet static energies can be derived 
along the same lines (cf. also the calculation of the US corrections for the $Q\bar Q$ octet potential 
in Ref.~\cite{Brambilla:1999xf}). 
In particular, in the equilateral limit one has to consider only the diagrams shown in Fig.~\ref{equimatch}, 
since octet-to-octet diagrams with an intermediate octet propagator in the loop are scaleless for $V_S^o=V_A^o=V^o$,
and thus vanish in dimensional regularization. Moreover, the US leading-order contribution for the symmetric octet is equal 
to the one for the antisymmetric octet; we call it, $\delta_{\rm US}^o$. The divergent parts of the diagrams shown 
in Fig.~\ref{equimatch} give rise to the following renormalization group equations valid for the singlet, 
octet and decuplet static potentials of three quarks located at the corners of an equilateral triangle of side length $r$:
\begin{equation}
\left\{
\begin{array}{l}
\displaystyle \mu\frac{\rm d}{{\rm d}\mu} V^s =-\frac{4}{3\pi}\alpha_{\rm s}r^2(V^o-V^s)^3+{\cal O}(\alpha_{\rm s}^5) \\
\displaystyle \mu\frac{\rm d}{{\rm d}\mu} V^o = \frac{1}{12\pi}\alpha_{\rm s}r^2\left[(V^o-V^s)^3+5(V^o-V^d)^3\right]+{\cal O}(\alpha_{\rm s}^5) \\
\displaystyle \mu\frac{\rm d}{{\rm d}\mu} V^d =-\frac{2}{3\pi}\alpha_{\rm s}r^2(V^o-V^d)^3+{\cal O}(\alpha_{\rm s}^5) \\
\displaystyle \mu\frac{\rm d}{{\rm d}\mu} \alpha_{\rm s} = \alpha_{\rm s}\beta(\alpha_{\rm s}) 
\end{array}
\right.
\,. 
\label{RG}
\end{equation}
The first equation is just the equilateral limit of Eq. \eqref{eq:RgVsgen}.
The last equation describes the running of the strong coupling constant, 
where $\beta(\alpha_{\rm s}) = - \alpha_{\rm s}\beta_0/(2\pi) + {\cal O}(\alpha_{\rm s}^2)$ is the beta function; 
the first coefficient of the beta function is $\beta_0 = 11 -2/3n_l$ with $n_l$ the number of light-quark flavors.  
By observing that
\begin{equation}
 V^o-V^s=-(V^o-V^d)+{\cal O}(\alpha_{\rm s}^3)\,,
\label{rel}
\end{equation}
as follows straightforwardly from the results of \cite{Brambilla:2009cd},
the system of equations (\ref{RG}) can be split into two sets of decoupled equations: 
\begin{equation}
\left\{
\begin{array}{l}
\displaystyle \mu\frac{\rm d}{{\rm d}\mu} V^s =-\frac{4}{3\pi}\alpha_{\rm s}r^2(V^o-V^s)^3+{\cal O}(\alpha_{\rm s}^5) \\
\displaystyle \mu\frac{\rm d}{{\rm d}\mu} V^o = -\frac{1}{3\pi}\alpha_{\rm s}r^2(V^o-V^s)^3+{\cal O}(\alpha_{\rm s}^5) \\
\displaystyle \mu\frac{\rm d}{{\rm d}\mu} \alpha_{\rm s} = \alpha_{\rm s}\beta(\alpha_{\rm s}) 
\end{array}
\right.
\,,
\label{RG2}
\end{equation}
and 
\begin{equation}
\left\{
\begin{array}{l}
\displaystyle \mu\frac{\rm d}{{\rm d}\mu} V^d =-\frac{2}{3\pi}\alpha_{\rm s}r^2(V^o-V^d)^3+{\cal O}(\alpha_{\rm s}^5) \\
\displaystyle \mu\frac{\rm d}{{\rm d}\mu} V^o = \frac{1}{3\pi}\alpha_{\rm s}r^2(V^o-V^d)^3+{\cal O}(\alpha_{\rm s}^5) \\ 
\displaystyle \mu\frac{\rm d}{{\rm d}\mu} \alpha_{\rm s} = \alpha_{\rm s}\beta(\alpha_{\rm s}) 
\end{array}
\right.
\,. 
\label{RG3}
\end{equation}
The two sets of equations can be solved as in~\cite{Pineda:2000gza} leading to\footnote{
All coupling constants in $V^s_{\rm NNLO}(r)$, $V^o_{\rm NNLO}(r)$ and $V^d_{\rm NNLO}(r)$ are evaluated at the scale $1/r$.
} 
\begin{eqnarray}
V^s(r;\mu) &=& V^s_{\rm NNLO}(r)-9\frac{\alpha_{\rm s}^3(1/r)}{\beta_0r} \ln\frac{\alpha_{\rm s}(1/r)}{\alpha_{\rm s}(\mu)}\,,
\label{vsr}\\
V^o(r;\mu)&=&V^o_{\rm NNLO}(r)-\frac{9}{4}\frac{\alpha_{\rm s}^3(1/r)}{\beta_0r}\ln\frac{\alpha_{\rm s}(1/r)}{\alpha_{\rm s}(\mu)}\,, 
\label{vor}\\
V^d(r;\mu) &=& V^d_{\rm NNLO}(r)+\frac{9}{2}\frac{\alpha_{\rm s}^3(1/r)}{\beta_0r}\ln\frac{\alpha_{\rm s}(1/r)}{\alpha_{\rm s}(\mu)}\,. 
\label{vdr}
\end{eqnarray}
The singlet static potential is known at NNLO, hence Eq.~\eqref{vsr} provides 
the complete expression of the singlet static potential at NNLL accuracy in an equilateral geometry.
This is the most accurate perturbative determination of this quantity. Instead neither the octet 
nor the decuplet potentials are known beyond NLO (see~\cite{Brambilla:2009cd}).

\section{Conclusions}
\label{sec:conclusions}
In the paper, we have reconsidered the construction of pNRQCD for systems 
made of three heavy quarks with equal masses. We have, in particular, rederived the pNRQCD 
Lagrangian in the static limit and put special attention to the symmetry under 
exchange of the heavy-quark fields. Although the symmetry is an obvious 
property of these systems, its consequences for the pNRQCD Lagrangian 
and in particular for its octet sector have been explored here for the first time.
Three static quarks may be cast either in a color-singlet, two distinct color-octets or 
a color-decuplet configuration. Whereas the color singlet is completely antisymmetric 
and the color decuplet is completely symmetric in the color-indices,
the color-octet transformations depend on the color indices that are exchanged.
The fact that color-octet fields are specially sensitive to the ordering of the quarks 
reflects in the fact that they mix, in general, under exchange of the heavy-quark fields
and dynamically through a one-gluon exchange.
As a consequence, also the octet potentials and the mixing potential transform non trivially under exchange symmetry;
we have listed their transformation properties in Eqs.~\eqref{Vrel1} and~\eqref{Vrel2}.

Thereafter, we have computed the leading ultrasoft contribution to the $QQQ$ 
singlet static energy, $\delta_{\rm US}^s$. Its expression can be found in~\eqref{US1}.
Because of the two different octet fields and their mixing, the calculation of $\delta_{\rm US}^s$ 
requires the evaluation of four diagrams and the resummation of the octet mixing potential 
for all of them. The calculation is therefore more involved than the analogous 
one of the US contribution in the $Q\bar{Q}$ case.
The expression for  $\delta_{\rm US}^s$ in the $QQQ$ case offers also a non-trivial 
test for the invariance under exchange symmetry; this has been performed in Sec.~\ref{sec:invdeltaUS}.
A consequence of the calculation of $\delta_{\rm US}^s$ at leading order is that we can determine 
the singlet static potential at order $\alpha_{\rm s}^4\ln\mu$, see Eq.~\eqref{Vs3loop}, 
and the singlet static energy at order $\alpha_{\rm s}^4\ln\alpha_{\rm s}$, see Eq.~\eqref{eq:E0full}.
These results represent the new computational outcome of this work and so far the most accurate determinations 
of the $QQQ$ singlet static potential and energy in perturbative QCD. 
The new contribution computed for the potential is valid for any configuration in space 
that the three quarks may take and it is a three-body interaction.
Together with the three-body interaction at two-loop order computed in~\cite{Brambilla:2009cd}
it may provide new insight on the emergence of a long-range three-body 
interaction governed by just one fundamental length that is observed in lattice studies 
(see e.g. \cite{Takahashi:2000te,Takahashi:2002bw,Takahashi:2004rw}).

In the last part of the paper, we have focused on the special situation
where the three quarks are located at the corners of an equilateral triangle of side
length $r$. In this limit, where the two octet potentials become degenerate,
we have solved the renormalization group equations for the color singlet, octet and decuplet 
potentials at NNLL accuracy. The corresponding expressions can be found 
in Eqs.~\eqref{vsr}-\eqref{vdr}. Hence, for an equilateral geometry, 
the $QQQ$ singlet static potential is now known up to order $\alpha_{\rm s}^{3}(\alpha_{\rm s}\ln\mu r)^n$  
for all $n \in \mathbb{N}_0$.

\begin{acknowledgments}
Work supported in part by DFG and NSFC (CRC 110), and by the DFG cluster of excellence
``Origin and structure of the universe'' (www.universe-cluster.de). 
F.K. gratefully acknowledges financial support from the FAZIT foundation and  inspiring discussions with E.~Thoma.
\end{acknowledgments}

\begin{appendix}

\section{Covariant derivative operators}
\label{app1}
In this appendix, we list the explicit matrix representations for the
covariant derivative operators in the octet and decuplet
representations of SU(3)$_c$ that appear in Eq.~(\ref{LpNRQCD1}).  
The SU(3)$_c$ covariant derivative is of the general form
\begin{align}
 D_{\mu}=\partial_{\mu}+igA^a_{\mu}T_r^a\,,
\end{align}
where $a=1, \ldots, 8$ and $T_r^a$ refers to the SU(3)$_c$ generators in the representation $r$.
The generators in the octet ($r=8$) and in the decuplet ($r=10$) representation are \cite{Brambilla:2005yk}
\begin{align}
 (T_8^a)_{bc}&= - if^{abc}\,,\hspace*{4cm}b,c=1, \ldots, 8, 
\nonumber\\
 (T_{10}^a)_{\delta\delta'}&= \frac{3}{2}\,\underline{\Delta}_{ijk}^{\delta}\lambda^a_{ii'}\underline{\Delta}_{i'jk}^{\delta'}\,, 
\hspace*{2.6cm}\delta,\delta'=1, \ldots, 10, 
\end{align}
where $f^{abc}$ are the structure constants of SU(3)$_c$.
An explicit representation of the decuplet tensor $\underline{\Delta}^{\delta}_{ijk}$ is in~(\ref{Delta}).

\section{The singlet static potential up to order $\alpha_{\rm s}^3$}
\label{app2}
We reproduce here for completeness the expression of the singlet static potential 
up to order $\alpha_{\rm s}^3$ computed in \cite{Brambilla:2009cd}:
\begin{eqnarray}
 V^s_{\rm NNLO}({\bf r}_1,{\bf r}_2,{\bf r}_3) &=& 
-\frac{2}{3}\sum_{i=1}^3\frac{\alpha_{\rm s}(1/|{\bf r}_i|)}{|{\bf r}_i|}
\left[1+\tilde{a}_1\frac{\alpha_{\rm s}(1/|{\bf r}_i|)}{4\pi}\right]
\label{Vs2loop}\\
&& \hspace{-22mm}
-\alpha_{\rm s}\left(\frac{\alpha_{\rm s}}{4\pi}\right)^2
\left[\frac{2}{3}\,\tilde{a}_{2,s}\left(\frac{1}{|{\bf r}_1|}+\frac{1}{|{\bf r}_2|}
+\frac{1}{|{\bf r}_3|}\right)+v_{\cal H}({\bf r}_2,{\bf r}_3)+v_{\cal H}({\bf r}_1,-{\bf r}_3)+v_{\cal H}(-{\bf r}_2,-{\bf r}_1)\right]\!.
\nonumber 
\end{eqnarray}
The one-loop and two-loop coefficients $\tilde{a}_1$ and $\tilde{a}_{2,s}$ 
depend on the number of light (massless) quark flavors, $n_l$, and are given by
\begin{eqnarray}
\tilde{a}_1&=&\frac{31}{3}+22\gamma_E-\left(\frac{10}{3}+4\gamma_E\right)\frac{n_l}{3}\,,
\\
\tilde{a}_{2,s}&=&\frac{4343}{18}+\frac{3\pi^4}{4}+\frac{121\pi^2}{3}+66\zeta(3)-484\gamma_E^2+204\gamma_E 
\nonumber\\
&& -\left(\frac{1229}{9}+\frac{44\pi^2}{3}+52\zeta(3)
-176\gamma_E^2+76\gamma_E\right)\frac{n_l}{3}+\left(\frac{100}{9}+\frac{4\pi^2}{3}-16\gamma_E^2\right)\left(\frac{n_l}{3}\right)^2 
\nonumber\\
&& +4\gamma_E\left(11-2\frac{n_l}{3}\right)\tilde{a}_1\,.
\end{eqnarray}
At two loop, a genuine three-body potential shows up. It is encoded in the function $v_{\cal H}$ defined as 
\begin{eqnarray}
v_{\cal H}({\bf r}_2,{\bf r}_3)&=& 
16\pi\int_0^1{\rm d}x \int_0^1{\rm d}y\,
\left\{\frac{\hat{\bf r}_2\cdot\hat{\bf r}_3}{|{\bf R}|}
\left[\left(1-\frac{M^2}{|{\bf R}|^2}\right)
\arctan\frac{|{\bf R}|}{M}+\frac{M}{|{\bf R}|}\right]\right. 
\\
&& \hspace*{3.6cm}
+\left.\frac{(\hat{\bf r}_2\cdot\hat{\bf R})(\hat{\bf r}_3\cdot\hat{\bf R})}{|{\bf R}|}
\left[\left(1+3\frac{M^2}{|{\bf R}|^2}\right)\arctan\frac{|{\bf R}|}{M}-3\frac{M}{|{\bf R}|}\right]\right\}, 
\nonumber
\end{eqnarray}
with ${\bf R}({\bf r}_2,{\bf r}_3)\equiv x{\bf r}_2-y{\bf r}_3$ and
$M({\bf r}_2,{\bf r}_3)\equiv |{\bf r}_2|\sqrt{x(1-x)}+|{\bf r}_3|\sqrt{y(1-y)}$.
Note that the three-body potential in (\ref{Vs2loop}) is manifestly invariant under the transformations \eqref{rwird1} and \eqref{rwird2}.
\end{appendix}

\end{document}